\documentclass[aps,reprint,nofootinbib,nobibnotes,amsmath,amssymb,floatfix]{revtex4-2}
\usepackage[T1]{fontenc}
\usepackage[utf8]{inputenc}
\usepackage[english]{babel}
\usepackage{graphicx}
\usepackage{mathtools}
\usepackage{bm}
\usepackage[dvipsnames]{xcolor}
\IfFileExists{orcidlink.sty}{\usepackage{orcidlink}}{\newcommand{\orcidlink}[1]{}}
\usepackage{slashed}
\usepackage{verbatim}
\usepackage[T1]{fontenc}
\usepackage{mathbbol}
\usepackage[dvipsnames]{xcolor}
\usepackage{orcidlink}

\usepackage{graphicx,color,overpic,mathtools}
\usepackage{amsthm,amsmath,amssymb,mathrsfs}
\usepackage{braket,bm,bbm,setspace}
\usepackage{cancel}
\usepackage{float}
\usepackage{xargs}
\definecolor{mypurple}{RGB}{128, 0, 128}
\usepackage{hyperref}
\hypersetup{
    colorlinks=true,
    linkcolor=mypurple, 
    citecolor=mypurple, 
    urlcolor=mypurple  
 }

\newcommand{\ed}{\mathop{}\!\mathrm{d}}
\newcommand{\viscgs}{\mathop{}\!\mathrm{g}/\mathrm{cm}/\mathrm{s}}
\newcommand{\epscgs}{\mathop{}\!\mathrm{g}/\mathrm{cm}^{3}}

\begin{document}

\title{Breakdown of hydrodynamic universality in causal first-order viscous neutron star oscillations}

\author{Lennox S. Keeble\,\orcidlink{0009-0009-5796-631X}} 
\email{keebls25@wfu.edu}
\affiliation{Department of Physics, Wake Forest University, Winston-Salem, North Carolina 27109, USA}
\author{Alejandro C\'ardenas-Avenda\~no\,\orcidlink{0000-0001-9528-1826}} 
\affiliation{Department of Physics, Wake Forest University, Winston-Salem, North Carolina 27109, USA}

\begin{abstract}
Hydrodynamic modeling of viscous neutron stars requires a choice of theory of relativistic hydrodynamics. That distinct formulations contain the same physical content in the long-wavelength limit is referred to as ``hydrodynamic universality.'' It has been recently shown that hydrodynamic universality can break down in radial neutron star oscillations: causality and finite-size effects restrict access to the long-wavelength universal hydrodynamic regime in sufficiently viscous stars. We show that this breakdown occurs for a broad range of causal hydrodynamic frames in Bemfica--Disconzi--Noronha--Kovtun (BDNK) theory. Increasing viscosity collectively pushes this class of causal frames away from the BDNK regime of validity, eventually forcing all to lie outside it at bulk viscosities in the range $\zeta\sim10^{30}$--$10^{31}\mathrm{g}/\mathrm{cm}/\mathrm{s}$. This breakdown of universality is accompanied by material changes in the oscillation mode structure: non-hydrodynamic modes become damped on timescales comparable to those of hydrodynamic modes, the first overtone can become longer lived than the fundamental mode, and unphysical instabilities with sub-millisecond timescales can emerge. Even in regimes where universality has broken down, we find that the inviscid threshold of gravitational collapse remains unchanged in BDNK theory. These results provide evidence that first-order theories of relativistic hydrodynamics may be insufficient for modeling sufficiently viscous neutron stars. 
\end{abstract}

\maketitle

\section{Introduction}
Neutron star oscillations encode information about the equation of state of the supra-nuclear matter in their interior~\cite{Andersson:1997rn, Kokkotas:1999bd, Benhar:2004xg}. The tidal excitation of non-radial modes in the inspiral phase of binary neutron star coalescences could leave observational imprints in the phase of the emitted gravitational-wave signal~\cite{Lai:1993di,Kokkotas:1995xe,Ho:1998hq,Hinderer:2016eia,Steinhoff:2016rfi,Pratten:2019sed}. Bulk viscous dissipation may play an important role in the inspiral and post-merger phases~\cite{Alford:2017rxf,Alford:2020lla,Most:2021zvc,Most:2022yhe,Ghosh:2023vrx, Ghosh:2025glz, Ghosh:2025wfx}, during which previous estimates suggest the bulk viscosity could reach $\zeta\sim10^{28}$--$10^{30}\viscgs$~\cite{Most:2021zvc,2001PhRvD..64h4003J,Alford:2020pld,Gusakov:2008hv}, with a bound of $\zeta\lesssim 10^{31}\viscgs$ from GW$170817$~\cite{Ripley:2023lsq}. Dissipative effects on neutron star tides~\cite{Bildsten:1992my,Kochanek:1992wk,Lai:1993di,Arras:2018fxj} and pulsations~\cite{Lindblom_1983,Cutler:1987,Sawyer:1989dp,Cutler:1990,Gusakov:2005dz} have long been studied, and may be accessible to third-generation ground-based gravitational-wave detectors~\cite{Ghosh:2023vrx, Ghosh:2025glz, Ghosh:2025wfx, Ripley:2023lsq, Ripley:2023qxo,Hammond:2022uua}, such as Cosmic Explorer and the Einstein Telescope~\cite{ET:2019dnz, ET:2025xjr, Reitze:2019iox}. These observational prospects have renewed interest in out-of-equilibrium effects in neutron star dynamics~\cite{Saketh:2024juq, Ripley:2023lsq, Ripley:2023qxo, HegadeKR:2024agt,Ghosh:2026ldg}, especially their modeling using causal and stable theories of relativistic hydrodynamics~\cite{Caballero_2025,Mendes:2025oib,Shum:2025jnl,Guerrieri:2026jbm,Denicol:2026qqk,Bussieres:2026rnz,Katagiri:2026jgp,Chabanov:2023abq, Chabanov:2023blf,Boyanov:2024,Redondo-Yuste:2025ktt,Redondo-Yuste:2024vdb,Keeble:2026bzo,Camelio:2022ljs,Camelio:2022fds}.

Hydrodynamics is an effective field theory (EFT)~\cite{Kovtun:2012rj} which integrates out degrees of freedom on microphysical length scales. Away from local thermodynamic equilibrium, only the stress-energy tensor $T^{ab}$ and current density $J^{a}$ retain unique microscopic definitions~\cite{Kovtun:2012rj,Kovtun:2019hdm}. This endows one with a choice of how to define the non-unique out-of-equilibrium hydrodynamic variables, in addition to the order of truncation in a gradient expansion of $T^{ab}$ and $J^{a}$. In other words, there is no single theory of dissipative relativistic hydrodynamics since there is no unique way in which to treat out-of-equilibrium degrees of freedom~\cite{Kovtun:2019hdm,Kovtun:2012rj,Hoult:2020eho,Rocha:2023ilf,disconzi_2024}. Nevertheless, different out-of-equilibrium formulations should agree in the near-equilibrium long-wavelength limit, which is captured by the Navier--Stokes constitutive relations. Following Ref.~\cite{Denicol:2026qqk}, we refer to this limit as the universal hydrodynamic regime.

Hydrodynamic modeling of neutron star matter requires a choice of theory of relativistic hydrodynamics. Of particular recent interest have been the first-order Bemfica--Disconzi--Noronha--Kovtun (BDNK)~\cite{Bemfica:2017wps, Bemfica:2019knx, Bemfica:2020zjp,Kovtun:2019hdm} and second-order Israel--Stewart (IS) theories~\cite{Muller_1967,ISRAEL1976213,ISRAEL1976310,ISRAEL1979341}. These theories are particularly appealing due to their causality and stability properties~\cite{Bemfica:2020zjp,Kovtun:2019hdm,Hiscock:1983,Olson_1990,Cordeiro_2025,Cordeiro:2025rkt,Cordeiro:2026jle,Bemfica_2019b,Pu:2009fj,Denicol:2008ha}. The extent to which these choices give equivalent physical predictions depends on whether the system under consideration can access the universal hydrodynamic regime~\cite{Denicol:2026qqk}. 

By comparing radial oscillation modes in causal IS theory and acausal relativistic Navier--Stokes, \textit{Denicol et al} recently showed that the universal hydrodynamic regime may not be reachable in sufficiently viscous neutron stars~\cite{Denicol:2026qqk}. Their finite size sets an upper limit on the wavelengths available to oscillation modes, while causality forces the microscopic IS relaxation time to grow with viscosity. For sufficiently large viscosities, the relaxation time is no longer well separated from the fundamental oscillation timescale, indicating a breakdown of the coarse-grained EFT regime.

In this work, we show that the breakdown of hydrodynamic universality in radial oscillations of sufficiently viscous stars occurs for a broad class of causal first-order hydrodynamic frames. Consistent with the results of Ref.~\cite{Denicol:2026qqk}, we find that this breakdown occurs at bulk viscosities $\zeta\sim10^{30}$--$10^{31}\viscgs$, towards the upper end of previous estimates and within the bound of $\zeta\lesssim 10^{31}\viscgs$ from GW$170817$. We also find that the inviscid threshold of gravitational collapse is unchanged in BDNK theory and provide a complementary analytical argument for this result.

The remainder of this work is structured as follows. In Sec.~\ref{sec:framework}, we summarize the perturbation equations and frequency-domain numerical methods used to solve them. We present analyses of the flat-space and radial neutron star mode spectra in Sec.~\ref{sec:results} and discuss our results in Sec.~\ref{sec:discussion}. In Appendices~\ref{app:flatspace} and~\ref{app:TD}, we provide further details about the flat-space derivation and consistency with the time-domain approach of Ref.~\cite{Keeble:2026bzo}. Throughout, we work in units in which $G=c=1$ and use abstract index notation.

\section{Perturbative framework}\label{sec:framework}
In this section, we summarize the perturbative and numerical methods we use to study radial oscillations of viscous stars. See, e.g., Refs.~\cite{Mendes:2025oib,Keeble:2026bzo} for a more detailed exposition. Below, we use $\varphi(t,r)$ as a metric function instead of $\nu(t,r)$ as used in Ref.~\cite{Keeble:2026bzo}.

\subsection{Background spacetime and constitutive relations}
We consider a spherically symmetric spacetime endowed with the Levi-Civita connection and metric
\begin{equation}
    \ed{s}^2 = -e^{\varphi(t,r)}\ed t^2+e^{\lambda(t,r)}\ed r^2+r^2\ed \Omega^2 \, \label{eq:metric}, 
\end{equation}
where $\ed\Omega^2$ is the unit round metric on $S^{2}$ and $r$ is the areal radius. The spacetime is coupled to a barotropic BDNK fluid at zero chemical potential without a conserved $U(1)$ charge, for which the constitutive relations are~\cite{Bemfica:2020zjp}
\begin{align}
 T^{ab}&=\left(\epsilon+\mathcal{A}\right)u^{a}u^{b}+(p+\Pi) \Delta^{ab}+2\mathcal{Q}^{(a} u^{b)}-2\eta\sigma^{ab},\nonumber\\
    \mathcal{A}&=\tau_{e}\left[u^{a}\nabla_{a}\epsilon+\left(\epsilon+p\right)\nabla_{a}u^{a}\right],\nonumber\\
    \Pi&=-\zeta\nabla_{a}u^{a}+\tau_{p}\left[u^{a}\nabla_{a}\epsilon+\left(\epsilon+p\right)\nabla_{a}u^{a}\right],\\
    \mathcal{Q}^{a}&=\tau_{q}\left[\left(\epsilon+p\right)u^{b}\nabla_{b}u^{a}+c_{s}^2\Delta^{ab}\nabla_{b}\epsilon\right],\nonumber\\
    \sigma^{ab}&=\frac{1}{2}\Delta^{ac}\Delta^{bd}\left(\nabla_{c}u_{d}+\nabla_{d}u_{c}-\frac{2}{3}\nabla_{e}u^{e}\Delta_{cd}\right),\nonumber
\end{align}
where $u^{a}$ is the timelike four-velocity of the fluid, $\Delta^{ab}=g^{ab}+u^{a}u^{b}$ is the projector orthogonal to $u^{a}$, $c_{s}^{2}\equiv\ed p /\ed\epsilon$, $\zeta$ and $\eta$ denote the bulk and shear viscosities, and $\tau_{e,p,q}$ are interpreted as relaxation times. We work in the zero-diffusion limit and consider a polytropic equation of state (EoS) $p(\epsilon)=\kappa \epsilon^{1+1/n}$ with $n=1$ and $\kappa=100\,\mathrm{km}^2$. We do not anticipate that these simplifying choices will affect our conclusions: Ref.~\cite{Guerrieri:2026jbm} found that finite-temperature thermal modes couple weakly to the fluid modes, while the results of Ref.~\cite{Denicol:2026qqk} were shown to hold for both a polytrope and a more realistic SLy EoS. Motivated by the expectation that bulk viscosity is the leading dissipative contribution in neutron stars~\cite{Ghosh:2025glz, Ghosh:2025wfx, Alford:2017rxf, Alford:2018lhf, Alford:2019qtm,Alford:2020lla, Most:2021zvc, Most:2022yhe,Lai:1993di}, we set $\eta=\zeta/10$. 

Setting $\tau_{e,p,q}=0$ recovers an instance of the class of Eckart~\cite{Eckart_1940} and Landau--Lifshitz~\cite{Landau_1987} theories which, in flat space, generically admit linearly unstable equilibrium states and superluminal Fourier modes~\cite{Hiscock_1985,Hiscock_1987}. We refer to the resulting theory as acausal relativistic Navier--Stokes. BDNK theory, in contrast, is a covariant generalization of the Navier--Stokes equations which is nonlinearly causal, linearly stable in flat space, locally well-posed, and strongly hyperbolic provided certain nonlinear constraints on the transport coefficients are satisfied~\cite{Bemfica:2020zjp}. The stability properties of BDNK theory can be attributed to its enforcing the second law of thermodynamics only for solutions in its regime of validity (hereafter referred to as the EFT regime)~\cite{Gavassino_2020}. The relaxation times $\tau_{e,p,q}$ function purely to regulate the behavior of the theory in the ultraviolet, rendering it causal. Their contribution to $T^{ab}$ is of second order on shell, since the out-of-equilibrium corrections satisfy $\mathcal{A}\sim\mathcal{O}(\partial^{2})$, $\Pi\sim -\zeta\nabla_{a}u^{a}+\mathcal{O}(\partial^{2})$, $\mathcal{Q}^{a}_{\rm{BDNK}}\sim\mathcal{Q}^{a}_{\rm{Eckart}}+\mathcal{O}(\partial^{2})$~\cite{Bemfica:2020zjp}. In the first-order EFT regime, different choices of $\tau_{e,p,q}$ therefore yield physically equivalent predictions.

The background spacetime is given by solutions of the Tolman--Oppenheimer--Volkoff (TOV) equations~\cite{Oppenheimer:1939ne,1939PhRv...55..364T},
\begin{subequations}
\begin{gather}
        \frac{\ed m}{\ed r}= 4 \pi  r^2 \epsilon,\quad \frac{\ed \varphi}{\ed r}= \frac{2 m+8 \pi  r^3 p}{r (r-2 m)},\\
        \frac{\ed p}{\ed r}= \frac{\left(m+4 \pi  r^3 p\right) (p+\epsilon)}{r (2 m-r)},
    \label{eq:TOV}
\end{gather}
\end{subequations}
which describe cold, spherically symmetric stars in hydrostatic equilibrium. The mass function $m(r)$ is defined by $\lambda=-\ln{(1-2m/r)}$. Besides the survey shown in Fig.~\ref{fig:compactness} and the near-collapse modes in Fig.~\ref{fig:modestruct}, we consider a background star with central density $\epsilon_{c}=4\times10^{15}\epscgs$, which has mass $M=1.326\,M_{\odot}$, radius $R=8.255\,\mathrm{km}$, and compactness $M/R=0.237$. 

\subsection{Numerical methods for the quadratic stellar eigenvalue problem}\label{ssec:algo}

We solve the BDNK radial perturbation equations in the frequency domain. We take as evolution equations for the Eulerian perturbation variables $\mathbf{U}(t,r)=\left[\delta{u},\delta\epsilon,\delta\lambda\right]^{\rm{T}}$, where $\delta{u}\equiv e^{\varphi/2}u^{r}$, the two second-order wave equations for $\delta{u}$ and $\delta\epsilon$ described in Appendix C of Ref.~\cite{Keeble:2026bzo}, hereafter referred to as equations $\textsf{A}$ and $\textsf{B}$, respectively. We also make use of the $tt$ and $tr$ Einstein equations (e.g., $G_{t}{}^{r}=8\pi T_{t}{}^{r}$), which are given by
\begin{subequations}
\begin{align}
    e^{-\lambda}\left[\frac{\lambda'}{r}\delta{\lambda}-\frac{\delta\lambda'}{r}-\frac{\delta\lambda}{r^2}\right]+8\pi\delta\epsilon+8\pi \delta\mathcal{A}&=0,\label{eq:tt}\\
    \partial_{t}\delta\lambda+8\pi re^{\lambda}\left[\left(\epsilon+p\right)\delta{u}+e^{\varphi/2}\delta\mathcal{Q}^{r}\right]&=0,\label{eq:tr}
\end{align}
\end{subequations}
where $\theta\equiv\nabla_{a}u^{a}$,
\begin{subequations}
\begin{align}
    \delta\theta&=e^{-\varphi/2}\left[\delta{u}'+\left(\frac{\lambda'}{2}+\frac{2}{r}\right)\delta{u}+\frac{1}{2}\partial_{t}\delta\lambda\right],\\
    \delta\mathcal{A}&=\tau_{e}\left[\left(\epsilon+p\right)\delta\theta+e^{-\varphi/2}\left(\partial_{t}{\delta\epsilon}+\epsilon'\delta{u}\right)\right].
\end{align}
\end{subequations}
We impose two boundary conditions (BCs) at the surface of the star $r=R$: $T^{tr}(R)\propto\mathcal{Q}^{r}(R)=0$ and $\Delta T^{r}{}_{r}(R)=0$, where $\Delta f(t,r)=f(t,r+\xi)-f(r)=\delta f+f'(r)\xi+\mathcal{O}(\xi^2)$ denotes the Lagrangian variation of a scalar function $f$. We impose the first BC by setting $\mathcal{Q}^{r}(R)=0$ in the $tr$ Einstein equation \eqref{eq:tr} at the stellar surface. The second BC is given explicitly by
\begin{align}
    \Delta T^{r}{}_{r}=c_{s}^2\delta\epsilon+p'\xi+\delta\Pi-2\eta\delta\sigma^{r}{}_{r}=0,
\end{align}
where the Lagrangian displacement $\xi=\delta{u}/(-i\omega)$,
\begin{subequations}
\begin{align}
    \delta\sigma^{r}{}_{r}&=\frac{2}{3}e^{-\varphi/2}\left[\frac{1}{2}\partial_{t}\delta\lambda+\partial_{r}\delta{u}+\left(\frac{\lambda'}{2}-\frac{1}{r}\right)\delta{u}\right],\\
    \delta\Pi&=\left[-\zeta+\tau_{p}\left(\epsilon+p\right)\right]\delta\theta+\tau_{p}e^{-\varphi/2}\left(\partial_{t}\delta\epsilon+\epsilon'\delta{u}\right).
\end{align}
\end{subequations}

Our numerical grid consists of $N+1$ points labeled by $j\in\{0,1,\ldots,N\}$ with $r_{j}=hj$, where the grid spacing $h=R/N$. The equations are discretized using second-order-accurate centered finite differences at interior points and one-sided differences at the boundaries. At the stellar center $j=0$, we impose the conditions $\delta{u}=\delta\lambda=\delta\epsilon'=0$, which are required for the perturbations to be regular. We solve equations $\textsf{A}$ and $\textsf{B}$ at interior points $j\in\{1,\ldots,N-1\}$. The third continuum equation we solve is the first-order $tt$ Einstein equation. We find it numerically advantageous to discretize this equation at the $N$ cell midpoints $r_{j-1/2}$, with $j\in\{1,\ldots,N\}$, utilizing a spatial Crank--Nicolson approach: second-order-accurate first-derivative stencils are centered at $r_{j-1/2}$ with step size $h/2$, coupling only neighboring grid points. Functions $f(r_{j-1/2})$ are replaced with $(f_{j}+f_{j-1})/2+\mathcal{O}(h^2)$, consistent with the second-order accuracy of our scheme. At the surface, we impose $\Delta T^{r}{}_{r}=0$ and the $tr$ Einstein equation with $\mathcal{Q}^{r}(R)=0$. The three equations at the center, $2(N-1)$ $\textsf{A}$ and $\textsf{B}$ equations, $N$ $tt$ Einstein equations, and two surface BCs give a total of $3(N+1)$ equations for $3(N+1)$ variables. We monitor the residuals of the unused third equation solved in Ref.~\cite{Keeble:2026bzo} and the $tt$ Einstein equation at interior points and the surface, respectively, checking that they converge to zero with resolution when evaluated on the discretized solutions obtained from our system of equations.

The spatial Crank--Nicolson approach is useful for obtaining convergent solutions to the eigensystem. The $tt$ equation is first order and hence has one linearly independent continuum solution. Using second-order accurate difference stencils centered on grid nodes couples three grid points, giving a second-order recurrence relation which adds an unphysical solution to the discretized equation that alternates in sign at successive nodes. This grid mode does not converge away with resolution, sometimes preventing eigenvalues from converging as the grid is refined. The spatial Crank--Nicolson approach couples only neighboring grid points, removing the grid mode and allowing for second-order convergence in the eigenvalue. 

We introduce a harmonic time dependence $\mathbf{U}(t,r)=e^{-i\omega{t}}\mathbf{U}(r)$ and solve the resulting equations in the frequency domain. The continuum equations we solve take the form
\begin{subequations}
\begin{align}
    \mathsf{A}&=\mathsf{A}^{(0)}-i\mu\omega_{\rm{PF}}\mathsf{A}^{(1)}-\mu^{2}\omega_{\rm{PF}}^2\mathsf{A}^{(2)},\\
    \mathsf{B}&=\mathsf{B}^{(0)}-i\mu\omega_{\rm{PF}}\mathsf{B}^{(1)}-\mu^{2}\omega_{\rm{PF}}^2\mathsf{B}^{(2)},\\
    (tt)&=(tt)^{(0)}-i\mu\omega_{\rm{PF}}(tt)^{(1)},
\end{align}\label{eq:continuum}
\end{subequations}
where $\mu\equiv\omega/\omega_{\rm{PF}}$ and $\omega_{\rm{PF}}$ denotes the corresponding perfect fluid eigenvalue. The $\mathsf{A}^{(k)}$ and $\mathsf{B}^{(k)}$ in Eq.~\eqref{eq:continuum} contain radial derivatives of $\mathbf{U}(r)$ up to order $2-k$,  with coefficients which are functions of the background, taken from Ref.~\cite{Keeble:2026bzo} for $\mathsf{A}$ and $\mathsf{B}$, and read off from Eq.~\eqref{eq:tt} for the $tt$ Einstein equation. The regularity conditions at the center have only $\mu^{0,1}$ contributions. The surface condition $\Delta T^{r}{}_{r}=0$ contains a $\mu^{-1}$ term due to the presence of $\xi=\delta{u}/(-i\omega)$. We therefore impose instead $\partial_{t}\Delta T^{r}{}_{r}=0$ which contains $\mu^{0,1,2}$ terms. The $tr$ equation at the surface contains only $\mu^{0,1}$ terms. The discretized system of equations can therefore be expressed in the form
\begin{align}
    Q(\mu)U\equiv\left(K+\mu C+\mu^{2}M\right)U=0,\label{eq:FDsystem}
\end{align}
where $K$, $C$, and $M$ are $3(N+1)\times3(N+1)$ matrices, with $K$ and $M$ real and $C$ imaginary. Given a solution $(\mu,U)$, $(-\bar{\mu},\bar{U})$ also solves the system, i.e., eigenvalues come in pairs reflected across the imaginary axis. 

We solve the quadratic eigensystem \eqref{eq:FDsystem} in two ways. The first method is a Newton iteration, which consists of iteratively solving the linearization of $Q(\mu+\Delta\mu)(U+\Delta U)=0$,
\begin{align}
    \begin{bmatrix}
    Q(\mu) & Q'(\mu)U\\
    e_{p}^{\rm{T}} & 0
    \end{bmatrix}
    \begin{bmatrix}
    \Delta U\\
    \Delta \mu
    \end{bmatrix}=-\begin{bmatrix}
    Q(\mu) U\\
    0
    \end{bmatrix},\label{eq:Newton}
\end{align}
for $\Delta U$ and $\Delta\mu$, setting $U\to U+\Delta{U}$ and $\mu\to\mu+\Delta\mu$, and repeating until $\left|{\Delta\mu}\right|<\max\left(1,|\mu|\right)\times 10^{-10}$ or $N_{\rm{steps}}$ Newton steps have been taken. Typically, $N_{\rm{steps}}=15$ is sufficient, although we sometimes go up to $N_{\rm{steps}}=60$ when $\hat\tau\leq10\hat{\tau}_{\min}$ (see their definitions in the next subsection). The second row in Eq.~\eqref{eq:Newton} corresponds to a choice of normalization of the eigenvector, $U_{p}=1$, where $p$ is the index of the largest element of the initial guess for the eigenvector. The second method consists of casting Eq.~\eqref{eq:FDsystem} as a generalized eigenvalue problem $Ax=\mu Bx$, where $V=\mu U$, $x=\left[U,V\right]^{\rm{T}}$,
\begin{align}
    A&=\begin{bmatrix}
    0 & I\\
    -K & -C
    \end{bmatrix},\quad 
    B=\begin{bmatrix}
    I & 0\\
    0 & M
    \end{bmatrix}.
\end{align}
The generalized eigenvalue problem is solved using the QZ algorithm~\cite{QZ} as implemented in the \texttt{Julia} programming language package \texttt{LinearAlgebra.jl}. 

Hydrodynamic modes are solved for numerically by a ``walk'' in viscosity for a fixed star and hydrodynamic frame. At the smallest viscosity, we use as an initial guess for the Newton algorithm the corresponding perfect fluid eigenmode. We use as an initial guess for larger viscosities the eigenmode at the previous viscosity in the walk. The mode spectrum becomes complicated at sufficiently large viscosities, with mode crossings and avoided crossings, where too large a step between successive viscosities in the walk can cause the tracked mode to jump to a distinct mode. We monitor these potential mode jumps by computing the overlap $\mathcal{O}(x,y)=|\langle x,y\rangle|/( \lVert x\rVert\,\lVert y\rVert)$ in $\xi$ between neighboring viscosities in the walk and the corresponding fractional change in the eigenvalue ($\langle\cdot,\cdot\rangle$ denotes the trapezoidal inner product and $\lVert x\rVert=\sqrt{\langle x,x\rangle}$). If the overlap is less than $0.9$ or the eigenvalue changes by more than $5\%$, we return to the previous step in the walk and halve the jump in viscosity. 

\begin{figure*}[hbt!]
    \centering
    \includegraphics[width=1.0\textwidth]{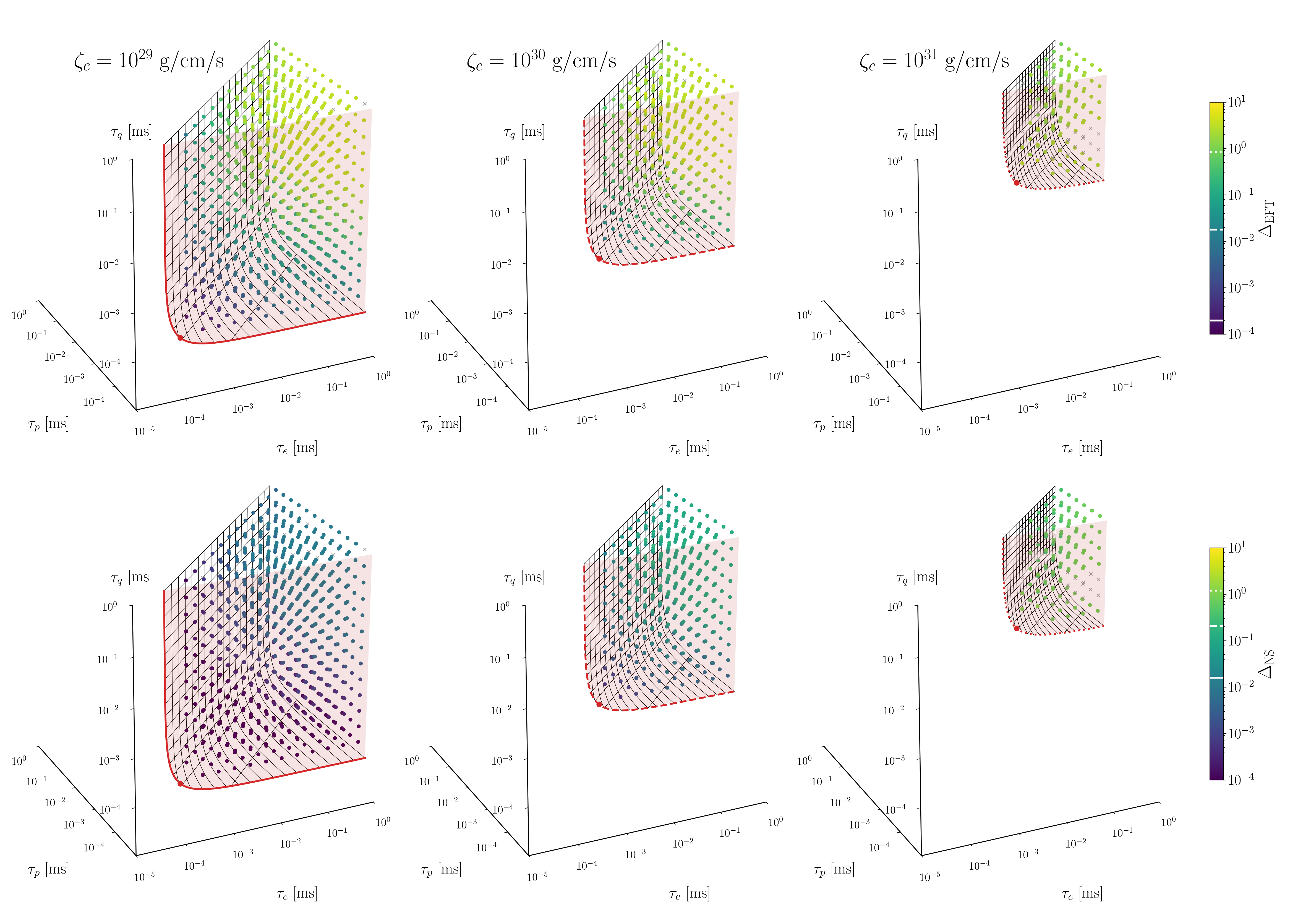}
    \caption{\textbf{Breakdown of hydrodynamic universality.} In each panel, the causal region of parameter space is bounded by the red and black surfaces, which correspond to the first and second conditions in Eq.~\eqref{eq:causalityb} satisfied with equality and evaluated at the center of the star. The fractional deviations $\Delta_{\rm{EFT}}$ and $\Delta_{\rm{NS}}$ are shown at each point (corresponding to a causal frame) in the upper and lower rows, respectively. From left to right, the columns correspond to central bulk viscosities $\zeta_{c}=10^{29,30,31}\viscgs$. Grey crosses denote frames for which the corresponding eigenvalue either did not converge or was not computed due to a mode at a lower viscosity in the walk becoming (unphysically) unstable. White lines on the colorbars, which ascend with viscosity, denote the minimum and maximum values, respectively, of $\Delta_{\rm{EFT}}$ and $\Delta_{\rm{NS}}$ for each viscosity. The causal region is pushed away from the origin as the viscosity is increased, eventually consisting only of frames which lie outside the EFT regime. Even for frames formally outside the EFT regime, the mode shift relative to acausal Navier--Stokes can remain small for sufficiently low viscosities.}
    \label{fig:universality}
\end{figure*}

To monitor convergence, the walk is performed separately on four grids with $N\in\{6401, 12801, 25601,51201\}$ (or from $12801$ to $102401$ for $\hat{\tau}\leq 10\hat{\tau}_{\min}$). An eigenvalue at a fixed viscosity is regarded as converged if (i) the absolute differences between successive grids shrink or are stably below $10^{-4}\times|\omega|$ on all grids, (ii) the last difference is less than $5\%$ of $|\omega|$, and (iii) the displacements on the finest two grids have overlap $>0.999$. We report the value on the second finest grid and store the error in the eigenvalue as the difference between the two finest grids. The modes shown in Figs.~\ref{fig:universality}--\ref{fig:compactness} were computed exclusively using the Newton walk in viscosity. The $52$ grey crosses in Fig.~\ref{fig:universality} correspond to frames and viscosities for which either the eigenvalue did not converge ($27$ in total), or the walk was terminated at a lower viscosity due to the tracked mode becoming unstable ($25$ in total, $21$ of which occur at $\zeta_{c}=10^{31}\viscgs$).

For the non-hydrodynamic modes shown in Fig.~\ref{fig:modestruct}, there is no a priori good initial guess for the eigenvalue and eigenvector. These modes are solved for using the QZ algorithm on grids with $N\in\{241,481,801\}$, keeping only eigenvalues with $\mathrm{Re}(\omega/2\pi)<10\,\rm{kHz}$ and $|\mathrm{Im}(\omega/2\pi)|<10\,\rm{kHz}$ which are robust against grid refinement. These eigenvalues are iterated further using Newton's method on grids with $N\in\{6401,25601,102401\}$.

\subsection{Causality conditions}
Sufficient conditions for the causal propagation of linear perturbations of the Einstein--BDNK system are given by~\cite{Keeble:2026bzo}
\begin{subequations}
    \begin{gather}
        \tau_{q},\tau_{p},\tau_{e} > 0 \, , \quad \eta,\zeta,\epsilon,p>0,\\
        \tau_{p}\geq \nu,\quad \tau_{q}\left(1-c_{s}^2-\frac{\tau_{p}}{\tau_{e}}\right)\geq \nu,\label{eq:causalityb}
        \end{gather}\label{eq:causality}
\end{subequations}
where $\nu\equiv \left(\zeta+4\eta/3\right)/\left(\epsilon+p\right)$. We confine our analysis to the region of parameter space in which the conditions~\eqref{eq:causality} hold, and refer to this as the causal region of parameter space. These conditions, however, were only shown to be sufficient, not necessary, so there may exist other classes of causal frames besides the one considered in this work.

\begin{figure*}[hbt!]
    \centering
    \includegraphics[width=1.0\textwidth]{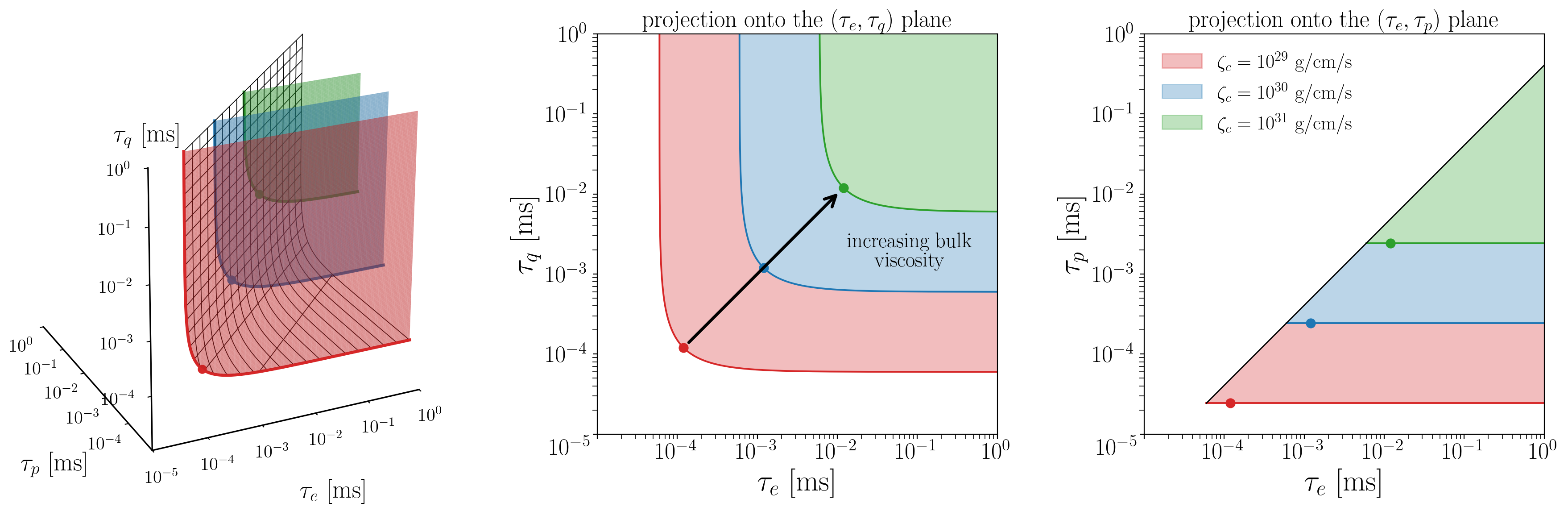}
    \caption{\textbf{Causal regions in parameter space.} In the left panel, we show boundaries of the regions satisfying the sufficient causality conditions~\eqref{eq:causality} for bulk viscosities $\zeta_{c}=10^{29,30,31}\viscgs$ in red, blue, and green, respectively. Here, all quantities entering Eq.~\eqref{eq:causality} are evaluated at the center of the star and the parameterization \eqref{eq:BDNK:Frame} is not used. The colored planes and black surface denote the first and second conditions, respectively, in Eq.~\eqref{eq:causalityb} satisfied with equality. We draw only one black surface since it moves negligibly for $\tau_{q}$ sufficiently large. The solid lines denote the curves along which the black and colored surfaces meet, with the solid circles denoting the points closest to the origin [through which passes the ``minimally causal'' ray~\eqref{eq:minray}]. The center and rightmost panels are projections of the causal regions in the first panel onto the $(\tau_{e},\tau_{q,p})$ planes. The red shading denotes the region in which only the smallest viscosity is causal, blue where the two smallest are causal, and green where all three are causal. Viscosity pushes these causal frames away from the origin and the EFT regime.}
    \label{fig:causality}
\end{figure*}

We show in Fig.~\ref{fig:causality} the region of causality in the three-dimensional parameter space $(\tau_{e},\tau_{p},\tau_{q})$ as defined by Eq.~\eqref{eq:causality} evaluated at the center of the star. Increasing bulk viscosity pushes the causal region away from the origin, at which acausal Navier--Stokes is located. We emphasize that Fig.~\ref{fig:causality} assumes only the stellar profile, and is therefore independent of the frame parameterizations introduced below, which are convenient for simplifying the causality conditions.

We consider frame parameterizations
\begin{gather}
    \begin{aligned}
    \eta = \hat{\eta}(\epsilon+p) Lc_{s}^{2},\quad&\zeta =\hat{\zeta}(\epsilon+p) L c_{s}^{2}\\
    \tau_{e}= \hat{\tau}_{e}L\hat{V},\quad\tau_{q}=\hat{\tau}_{q}&L\hat{V},\quad \tau_{p}=\hat{\tau}_{p} L\hat{V} c_{s}^{2},\\
    \hat{\tau}_{e}\equiv\hat{\tau},\quad \hat{\tau}_{p}=&\alpha\hat{\tau},\quad\hat{\tau}_{q}=\beta \hat{\tau},
    \end{aligned}\label{eq:BDNK:Frame}
\end{gather}
where $\alpha,\beta>0$, $\hat{V}\equiv\hat{\zeta}+4\hat{\eta}/3$, and we take $L=1\,\mathrm{km}$. Throughout, we specify the value of the central bulk viscosity, $\zeta_{c}\equiv\zeta(r=0)$, which fixes the value of $\hat{\zeta}$. A fixed pair $(\alpha,\beta)$ determines a ray in the three-dimensional parameter space $(\tau_{e},\tau_{p},\tau_{q})$ parameterized by $\hat{\tau}$. Along a ray, the causality conditions \eqref{eq:causalityb} become
\begin{equation}
    \hat{\tau}\geq 1/\alpha \, , \quad c_s^2 \leq c^{2}_{\mathrm{max}}(\hat{\tau};\alpha,\beta)\equiv\frac{\beta \hat{\tau}}{1+\beta \hat{\tau}\left(1+\alpha\right)}.\label{eq:Causality}
\end{equation}
From Eq.~\eqref{eq:Causality}, it follows that $\partial_{\hat{\tau}}(c_{\max}^2)=\beta\left[1+\beta\hat{\tau}\left(1+\alpha\right)\right]^{-2}>0$. Hence, if a frame is causal at some $\hat{\tau}_{0}$, then it is causal for all $\hat{\tau}>\hat{\tau}_{0}$---causality only places a lower bound on $\hat{\tau}$. Also, $\lim_{\hat{\tau}\to\infty}c_{\max}^2=\left(1+\alpha\right)^{-1}$, so $\alpha$ controls the maximum attainable sound speed, which approaches unity only if $\alpha\to0$. Since $c_{s}^{2}$ is a monotonically decreasing function of $r$, causality requires
\begin{align}
    \hat{\tau}\geq\hat{\tau}_{\min}\equiv\max\left\{\frac{1}{\alpha}, \frac{c_{0}^2}{\beta\left[1-c_{0}^{2}\left(1+\alpha\right)\right]}\right\},\,\alpha<\frac{1-c_{0}^2}{c_{0}^2},\label{eq:Causality2}
\end{align}
where $c_{0}^{2}\equiv c_{s}^{2}(r=0)$. We define  the minimally causal ray,
\begin{align}
    \alpha=\frac{1-c_{0}^2}{2c_{0}^2}, \quad\beta=1,\label{eq:minray}    
\end{align}
which includes the frame closest to the origin $\tau_{e,p,q}=0$ [located at $\hat{\tau}=2c_{0}^2/(1-c_{0}^2)$; see Fig.~\ref{fig:causality}].

Causality requires a minimum value of $\hat{\tau}$, and, from Eq.~\eqref{eq:BDNK:Frame}, a minimum value of the dimensionful $\tau_{i}$ which increases with viscosity. This interplay between causality and viscosity can be qualitatively understood from non-relativistic fluid dynamics. Over a time $t$, viscosity diffuses momentum over a length scale $\sqrt{\nu{t}}$, where $\nu$ is the kinematic viscosity. Increasing viscosity diffuses momentum more aggressively, thereby requiring a longer diffusion timescale for causality to hold. Although the diffusion equation is parabolic with infinite propagation speeds, its hyperbolic completion given by the telegrapher's equation has characteristic speeds $\sqrt{\nu/\tau}$, for which this intuition is more precise.

The validity of first-order hydrodynamics, on the other hand, requires the relaxation times not to be too large, for otherwise the solution depends on the causal regulators $\tau_{i}$. The first-order EFT regime therefore schematically corresponds to a window $\hat{\tau}\in(\hat{\tau}_{\min},\hat{\tau}_{\max})$, where $\hat{\tau}_{\min}$ is set by causality and $\hat{\tau}_{\max}$ set by the validity of the EFT assumptions.

\section{Breakdown of first-order hydrodynamic universality}\label{sec:results}
\subsection{Flat-space analysis}
Perturbations of infinite homogeneous fluids in flat spacetime capture qualitative features of the discrete stellar mode spectrum remarkably well~\cite{Mendes:2025oib,Guerrieri:2026jbm}. We first perform such an analysis to understand the effects of causality, viscosity, and finite size in a simplified setting. 

In the rest frame of the background fluid, the sound channel consists of two hydrodynamic modes~\cite{Kovtun:2019hdm},
\begin{align}
    \omega_{s}(k)&=c_{s}k-\frac{i\nu}{2}k^2-\frac{\nu^{2}}{8c_{s}}k^{3}\nonumber\\
    &\quad-\frac{i\nu}{2}\tau_{q}\left(\nu-\tau_{p}+c_{s}^{2}\tau_{e}\right)k^4+\mathcal{O}(k^{5}),
\end{align}
and two non-hydrodynamic modes $\omega_{e,q}=-i/\tau_{e,q}+\mathcal{O}(k^{2})$, where $k$ is the magnitude of the spatial wave vector and the other sound mode is given by $\omega_{s}(-k)$. Defining $\mathrm{Im}(\omega_{s})\equiv-\left(\Gamma_{\rm{NS}}+\delta\Gamma\right)+\mathcal{O}(k^5)$, we have
\begin{figure*}[hbt!]
    \centering
    \includegraphics[width=1.0\textwidth]{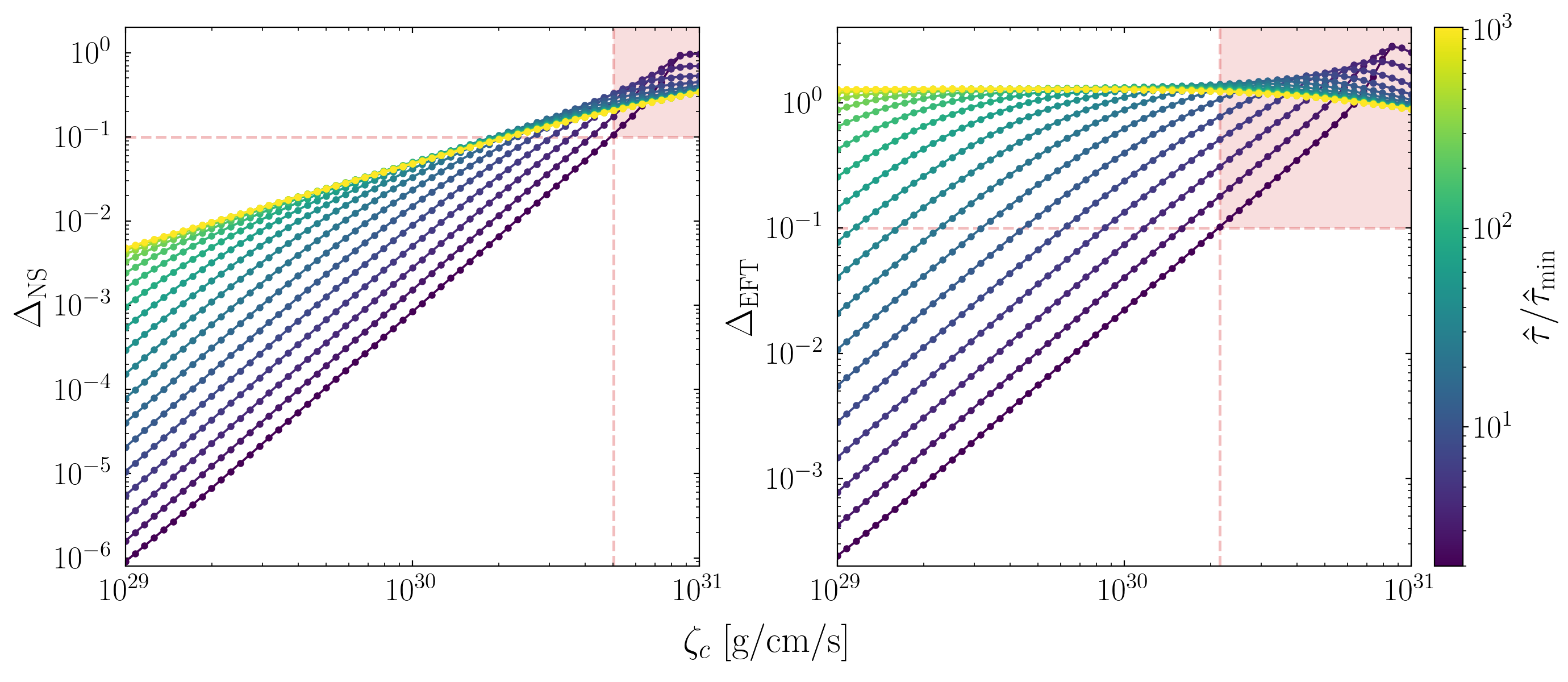}
    \caption{\textbf{Viscosity dependence of universality breakdown.} In the left and right panels, respectively, $\Delta_{\rm{NS}}$ and $\Delta_{\rm{EFT}}$ are evaluated along the minimally causal ray~\eqref{eq:minray} functions of the central bulk viscosity and choice of $\hat{\tau}$ (see colorbar). We shade in red the viscosities for which all considered frames have $\Delta_{\rm{NS,EFT}}>0.1$. As the viscosity increases, the range of values the fractional deviations take on narrows and both approach unity, indicating the breakdown of hydrodynamic universality.}
    \label{fig:ray}
\end{figure*}
\begin{align}
    \Delta\equiv\frac{\delta\Gamma}{\Gamma_{\rm{NS}}}&=\tau_{q}\left(\nu+c_{s}^2\tau_{e}-\tau_{p}\right)k^2.\label{eq:MinkFrameCorrection}
\end{align}
As shown in Appendix~\ref{app:flatspace}, imposing the sufficient conditions for the causal propagation of linear stellar perturbations~\eqref{eq:causality}, we find that $\Delta\geq\Delta_{\min}$ for viscosities $\zeta\gtrsim\zeta^{*}$, where
\begin{align}
     \zeta^{*}\sim\sqrt{\Delta_{\min}}\frac{1-c_{s}^2}{2c_{s}}\frac{\epsilon+p}{k},\label{eq:estimate}
\end{align}
provided $c_{s}^{2}\geq1/2$. Throughout, we take as a diagnostic of the breakdown of hydrodynamic universality the fractional frame deviation $\Delta$ being at least $10\%$ for all causal frames, i.e., $\Delta_{\min}=0.1$. Taking the wave number to be $k\sim\pi/R$, where $R$ is the radius of the star, we find
\begin{align}
    \zeta^*\sim\frac{9\left(1-{c}_{s}^2\right)}{2\pi {c}_{s}}\times 10^{30}\,\viscgs,
\end{align}
where we have taken $\epsilon+p\sim10^{15}\,\mathrm{g/cm^3}$ and $R\sim10\,\rm{km}$. For the background star considered here, Eq.~\eqref{eq:estimate} gives $\zeta^*\sim 3\times 10^{30}\,\viscgs$, which is close to the value of $\zeta^*\sim 2\times 10^{30}\,\viscgs$ from Fig.~\ref{fig:ray} below. 

The effects of finite size discussed in Ref.~\cite{Denicol:2026qqk} manifest themselves clearly in Eqs.~(\ref{eq:MinkFrameCorrection}--\ref{eq:estimate}). If the system can access the universal regime $k\to0$, $\Delta$ can be made arbitrarily small and $\zeta^{*}$ arbitrarily large.

\subsection{Neutron star oscillations}

For the viscosities we consider, $\zeta_{c}=10^{29}$--$10^{31}\viscgs$, it is possible that at the upper range there are order-unity corrections to the energy-momentum tensor which would necessarily place the theory outside the EFT regime. To monitor this, we decompose the BDNK frequencies as
\begin{align}
    \omega_{\rm{BDNK}}&=\omega_{\rm{PF}}+\underbrace{\left(\omega_{\rm{NS}}-\omega_{\rm{PF}}\right)}_{\text{viscous effects}}+\underbrace{\left(\omega_{\rm{BDNK}}-\omega_{\rm{NS}}\right)}_{\text{frame effects}},\label{eq:decomp}
\end{align}
and use the condition $\Delta_{\rm{GE}}\equiv\left|1-\omega_{\rm{NS}}/\omega_{\rm{PF}}\right|\ll1$ as a proxy for the validity of the first-order truncation of the gradient expansion. The deviation $\Delta_{\rm{GE}}$ scales linearly with $\zeta$~\cite{Keeble:2026bzo}, and, for the background star with $\epsilon_{c}=4\times10^{15}\epscgs$ considered in this work, ranges from $0.0038$ to $0.38$ across $\zeta_{c}=10^{29}$--$10^{31}\viscgs$. While this proxy remains under control for these viscosities, it begins to approach order unity for the largest viscosity, suggesting that the first-order truncation may not be sufficient for treating larger viscosities. Importantly, $\Delta_{\rm{GE}}\approx0.08$ at $\zeta_{c}=2\times10^{30}\viscgs$, the threshold above which all frames on the minimally causal ray have $\Delta_{\rm{EFT}}>0.1$ (see below), so gradient expansion effects and frame effects remain well separated there.

The effect of transforming between first-order hydrodynamic frames enters in Eq.~\eqref{eq:decomp} through the ratio of the last two terms. We define the fractional deviations
\begin{align}
    \Delta_{\rm{EFT}}=\left|\frac{\omega_{\rm{BDNK}}-\omega_{\rm{NS}}}{\omega_{\rm{NS}}-\omega_{\mathrm{PF}}}\right|,\
    \Delta_{\rm{NS}}=\left|\frac{\omega_{\rm{BDNK}}-\omega_{\rm{NS}}}{\omega_{\rm{NS}}}\right|,\label{eq:ΔNS}
\end{align}
such that the EFT regime corresponds to $\Delta_{\rm{EFT}}\ll1$: the change in the fundamental mode frequency upon transforming to a causal frame from acausal Navier--Stokes should be small relative to the change due to including viscosity~\cite{Bea2023}. Assuming the first-order truncation is valid from the perspective of the gradient expansion, the theory can exit the first-order EFT regime due to the breakdown of universality. The deviation $\Delta_{\rm{EFT}}$ assesses first-order frame robustness, a lack of which indicates that the system is not in the universal hydrodynamic regime. We also keep track of $\Delta_{\rm{NS}}$, which encodes the effect of transforming to a causal frame from acausal Navier--Stokes.

We show in Fig.~\ref{fig:universality} the fractional deviations $\Delta_{\rm{NS,EFT}}$ over a broad range of hydrodynamic frames filling the causal region in parameter space defined by Eq.~\eqref{eq:causality}. The causal region for the $\zeta_{c}=10^{29}\viscgs$ star admits frames with sufficiently small $\tau_{i}$'s such that the theory remains in the first-order EFT regime. Increasing viscosity pushes the causal region of parameter space out to larger $\tau_{i}$. For $\zeta_{c}=10^{31}\viscgs$, all the causal frames lie outside the EFT regime (i.e., $\Delta_{\rm{EFT}}\geq0.85$).

\begin{figure*}[hbt!]
    \centering
    \includegraphics[width=1.0\textwidth]{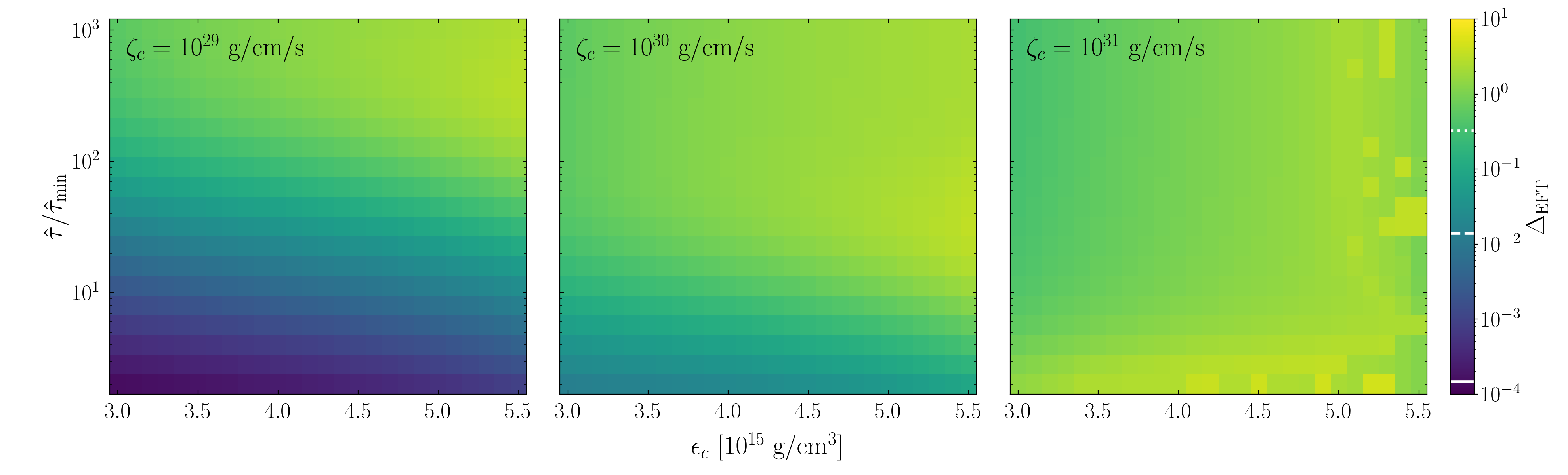}
    \caption{\textbf{Central density dependence of universality breakdown.} In each panel, $\Delta_{\rm{EFT}}$ (see colorbar) is shown as a function of $\hat{\tau}$ along the minimally causal ray~\eqref{eq:minray} for central densities in the range $\epsilon_{c}\in[3,5.5]\times10^{15}\epscgs$. All other results in the main text, besides the threshold of collapse, set $\epsilon_{c}=4\times10^{15}\epscgs$. The central bulk viscosity is fixed to $\zeta_{c}=10^{29,30,31}\viscgs$ from left to right, respectively. The white lines on the colorbar denote the minimum value of $\Delta_{\rm{EFT}}$ in each panel, which increases with $\zeta_{c}$. For the largest viscosity, the fundamental mode eventually migrates to the imaginary axis, where it collides with its reflected partner mode and splits into two different purely damped roots. Which root to follow becomes ambiguous, and the different central densities and frames are not guaranteed to follow the same root, giving some local cell-to-cell variation in the rightmost panel.
    With increasing viscosity, the minimum attainable value of $\Delta_{\rm{EFT}}$ by transforming to a causal frame increases. Along the minimally causal ray, this breakdown of universality does not depend strongly on the compactness of the star.}
    \label{fig:compactness}
\end{figure*}

The lower row of Fig.~\ref{fig:universality} demonstrates that the effects of formally leaving the EFT regime can be suppressed by small viscosities. For $\zeta_{c}=10^{29}\viscgs$, frames outside the EFT regime still give a fundamental mode close to that of acausal Navier--Stokes (i.e., $\Delta_{\rm{NS}}\leq0.016$). This is also apparent in Fig.~\ref{fig:ray}, in which we show $\Delta_{\rm{NS,EFT}}$ along the minimally causal ray~\eqref{eq:minray}. For $\zeta_{c}=10^{29}\viscgs$, $\Delta_{\rm{EFT}}$ reaches up to order unity as $\hat{\tau}$ is increased, while $\Delta_{\rm{NS}}<0.01$. As viscosity increases, the ranges of $\Delta_{\rm{NS,EFT}}$ across frames narrow, and both approach order unity. This is the breakdown of universality---for example, for $\zeta_{c}\gtrsim2\times 10^{30}\viscgs$, there does not exist a causal frame along the minimally causal ray for which $\Delta_{\rm{EFT}}<0.1$. Although we predominantly consider the case of a star with central density $\epsilon_{c}=4\times10^{15}\epscgs$, we show in Fig.~\ref{fig:compactness} that the shrinking of the window $\hat{\tau}\in\left(\hat{\tau}_{\min},\hat{\tau}_{\max}\right)$, which results in the breakdown of universality, does not depend strongly on the compactness of the star.

\begin{figure*}
\includegraphics[width=\textwidth]{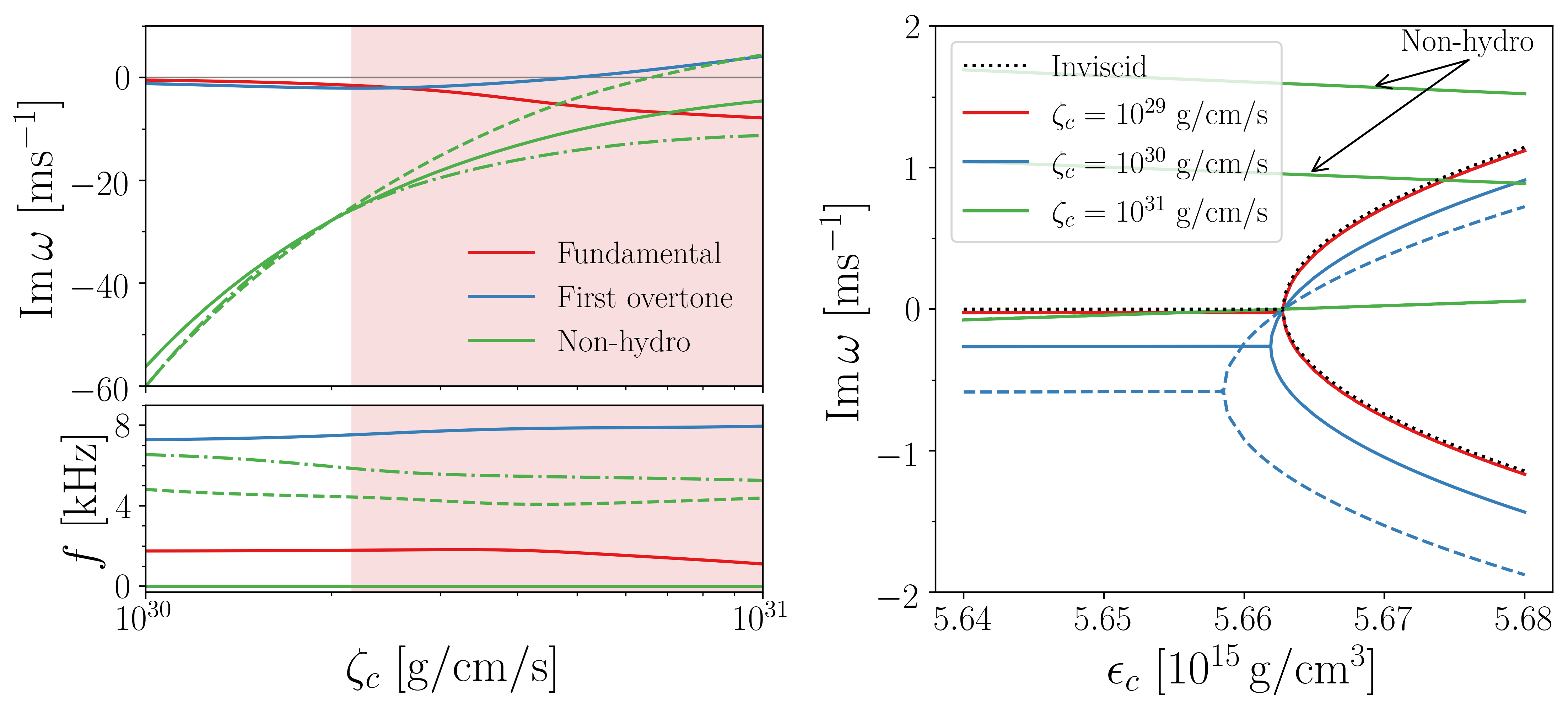}
 \caption{\textbf{Mode structure near the breakdown of universality and the threshold of gravitational collapse.} Left: fundamental mode, first overtone, and first few non-hydrodynamic modes along the minimally causal ray~\eqref{eq:minray} with $\hat{\tau}/\hat{\tau}_{\min}=8$, $\epsilon_{c}=4\times10^{15}\epscgs$, and $2\pi f\equiv\mathrm{Re}(\omega)$. The shaded region is where every frame on the minimally causal ray has $\Delta_{\rm{EFT}}>0.1$ (also shaded red in Fig.~\ref{fig:ray}). In this region, the damping rates of the non-hydrodynamic modes become comparable to those of the hydrodynamic modes and the first overtone becomes longer lived than the fundamental mode, eventually becoming unstable. Right: fundamental mode for the inviscid case and three central bulk viscosities (see legend) with $\hat{\tau}/\hat{\tau}_{\min}=2$, and non-hydrodynamic modes for the largest viscosity which are unstable before the perfect-fluid collapse threshold. The dashed lines for the two smaller viscosities have $\hat{\tau}/\hat{\tau}_{\min}=8$, which overlap with the solid line for $\zeta_{c}=10^{29}\viscgs$ but are clearly distinguishable for $\zeta_{c}=10^{30}\viscgs$, indicating an unphysical frame dependence. Nevertheless, the central density at which the fundamental mode goes unstable remains unchanged for all three viscosities and frames relative to the inviscid case.}\label{fig:modestruct}
\end{figure*}

For the background star considered in Fig.~\ref{fig:universality}, which is below the perfect-fluid threshold $\epsilon^{*}_{c}\approx 5.663\times10^{15}\epscgs$~\cite{Keeble:2026bzo} of gravitational collapse, there are unphysical BDNK mode instabilities outside the EFT regime. In particular, $25$ of the grey crosses in Fig.~\ref{fig:universality} correspond to eigenvalues for which the walk in viscosity was terminated due to the tracked mode becoming unstable ($21$ of which occur at $\zeta_{c}=10^{31}\viscgs$). Such unphysical instabilities are also manifest in the left panel of Fig.~\ref{fig:modestruct}, in which we show the mode structure near the breakdown of universality. The damping rate of the fundamental mode increases, while those of the first overtone and non-hydrodynamic modes decrease. The first overtone and one of the non-hydrodynamic modes eventually become unstable with timescales $\sim0.2\,\mathrm{ms}$ at $\zeta_{c}=10^{31}\viscgs$. These results provide evidence that, although linearly stable in flat space~\cite{Bemfica:2020zjp,Kovtun:2019hdm} and in zero-diffusion radial oscillations in a small-viscosity expansion~\cite{Caballero_2025}, BDNK theory can admit unphysical instabilities, in this case occurring as hydrodynamic universality breaks down at sufficiently large viscosities.

The right panel of Fig.~\ref{fig:modestruct} shows the mode structure near the threshold of gravitational collapse. As $\mathrm{Re}(\omega)\to0$, the two mirror modes with oscillation frequency $\pm|\mathrm{Re}(\omega)|$ approach each other, colliding on the imaginary axis and splitting into two purely imaginary eigenvalues. For a perfect fluid, this occurs precisely at the threshold of collapse. With increasing viscosity, the collision occurs at smaller central densities and outside the range of Fig.~\ref{fig:modestruct} for the largest viscosity. The manifest frame dependence for $\zeta_{c}=10^{30}\viscgs$ and the unphysical non-hydrodynamic instabilities for $\zeta_{c}=10^{31}\viscgs$ indicate the theory is outside the EFT regime. Nevertheless, the central density at which the fundamental mode becomes unstable, i.e., the threshold of gravitational collapse, remains unchanged. This is expected since, for the constitutive relations considered here, the BDNK mode spectrum contains the zero mode of the perfect fluid spectrum. For a static perturbation $\omega=\delta{u}=0$, the dissipative corrections $\mathcal{A}=\Pi=\sigma^{ab}=0$ and, by the TOV equations, $\mathcal{Q}^{a}=0$. The perturbation equations reduce to those of a perfect fluid, which admit a zero-frequency mode only at the threshold of gravitational collapse, beyond which the fundamental mode is unstable~\cite{Kokkotas:2000up,Bardeen:1966}. 

The time-domain simulations of Ref.~\cite{Keeble:2026bzo} reported a slight shift in the inviscid threshold of gravitational collapse within BDNK theory. This shift likely resulted from the backwards differences at the stellar surface, initial data built from acausal Navier--Stokes eigenvectors (which excite distinct modes), a numerically unstable grid mode, or a combination of these factors. The frequency-domain methods employed in this work allow us to more accurately investigate the mode structure near this threshold by solving directly for the eigenvalues and eigenvectors of the BDNK system, and employing physical boundary conditions at the surface which are able to preserve the location of the zero-frequency mode which, as argued above, is shared with the inviscid spectrum.

\section{Discussion}\label{sec:discussion}
We have shown that causality, viscosity, and finite-size effects prevent causal first-order radial neutron star oscillations from reaching the universal hydrodynamic regime for bulk viscosities $\zeta\sim10^{30}$--$10^{31}\viscgs$. This breakdown of hydrodynamic universality occurs for a broad class of causal first-order frames, providing evidence that first-order theories may be insufficient for hydrodynamic modeling of sufficiently viscous neutron stars. As first shown in Ref.~\cite{Denicol:2026qqk}, this breakdown emerges due to the fundamental ingredients of causality and finite size, extending beyond the specific case of radial neutron star oscillations considered here.

Properties of the first-order mode spectrum---the first overtone becoming the longest lived and the emergence of dynamically important non-hydrodynamic modes---are consistent with the IS mode spectrum studied in Ref.~\cite{Denicol:2026qqk}. Our results provide evidence that, at least at first order, such features occur in a regime in which universality has broken down. A systematic second-order causal-frame analysis would determine whether universality is restored at these viscosities and such features are physical, or if they also occur outside the second-order EFT regime, in which case a far-from equilibrium prescription may be necessary. The second-order theory presented in Ref.~\cite{Noronha:2021syv} would be well suited for such a study, since its first-order truncation is BDNK theory and it includes IS theory as a special case.

Throughout, we have used acausal Navier--Stokes (more generally, Eckart's theory) as a physical benchmark despite its pathologies, since BDNK theory contains the same physical content on-shell~\cite{Bemfica:2020zjp}. In the EFT regime, this physical content should be inherited by the eigenvalues since they are properties of the viscous star which follow from the constitutive relations. However, in astrophysical settings, hydrodynamic mode universality does not imply that models using different hydrodynamic theories will agree, as the modes alone do not determine the dynamics of the system. How they are excited, i.e., initial data, and the subsequent evolution of the system, as governed by the principal part of the partial differential equations of motions and the entropy current, will also determine the long-term dynamics. The latter two components differ substantially between Eckart's theory and BDNK: the former is parabolic and exactly enforces the second law of thermodynamics~\cite{Eckart_1940,Hiscock_1987}, while the latter is hyperbolic and enforces the second law only on solutions in its regime of validity~\cite{Bemfica:2020zjp,Kovtun:2019hdm}. Even in the lower-viscosity regimes in which hydrodynamic mode universality appears to hold, a detailed comparison of time-domain evolutions and long-term dynamics between different first-order theories is necessary to establish whether a more general, physically-realizable universal hydrodynamic regime exists. We leave such considerations to future work.

The focus in this work on causal first-order oscillations is motivated by the fact that causality is an essential ingredient for the general stability properties of dissipative field equations~\cite{Bemfica:2020zjp,Gavassio_2022:1,Gavassino_2022:2}. The zero-diffusion acausal Navier--Stokes theory considered in this work is linearly stable in flat space in the rest frame of the background fluid~\cite{Hiscock_1985}, and in radial oscillations of neutron stars~\cite{Lindblom_1983,Caballero_2025}. However, in Minkowski spacetime, a dissipative theory that is stable in one frame of reference is stable in all reference frames related by a Lorentz transformation if and only if it is causal~\cite{Gavassio_2022:1,Bemfica:2020zjp}. Indeed, all first-order theories belonging to the general class considered in Ref.~\cite{Hiscock_1985} are unstable in a boosted frame.

In numerical relativity, more general reference frames, and transformations between frames, are often used. Such simulations with matter prescribed within first-order hydrodynamics therefore require causal and stable formulations such as BDNK. The results of this work, however, provide evidence that BDNK theory can develop unphysical mode instabilities in astrophysical settings. In particular, the breakdown of universality in sufficiently viscous, gravitationally stable neutron stars can lead to causal BDNK frames developing unphysical modes that grow on sub-millisecond timescales. Were they to manifest, these instabilities would cause numerical-relativity simulations to exhibit unphysical blow-ups, which is especially relevant given the proliferation of general-relativistic dissipative hydrodynamic simulations~\cite{Chabanov:2023abq, Chabanov:2023blf,Shum:2025jnl,Lier:2026owd,Most:2026uga}.

\section*{Acknowledgments} 
We thank Fabio Bemfica, Daniel Caballero, Marcelo Disconzi, Raissa Mendes, and Jorge Noronha for helpful comments and discussions. We thank Wolfgang Br\"oder for useful discussions about frequency-domain solvers. Anthropic Claude Fable $5.1$ was used to accelerate numerical implementations, data post-processing, and manuscript proofreading. All scientific analyses and conclusions were produced by the authors, who take full responsibility for the content of this work. Computations were performed using the Wake Forest University (WFU) High Performance Computing Facility, a centrally managed computational resource available to WFU researchers, including faculty, staff, students, and collaborators~\cite{WakeHPC}.

\begin{figure*}[hbt!]
    \centering
    \includegraphics[width=\textwidth]{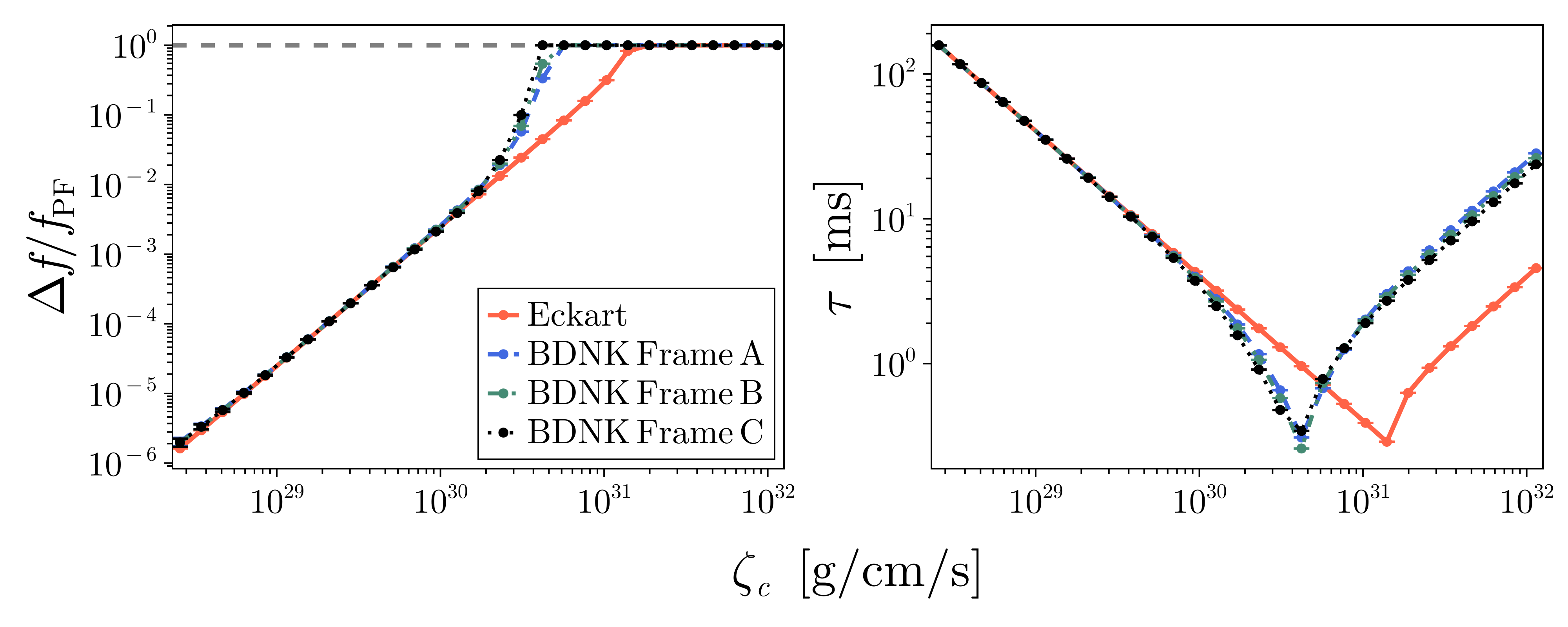}
    \caption{Left: fractional frequency shifts $\Delta{f}/f_{\rm{PF}}$, where $\Delta{f}=f_{\mathrm{PF}}-f_{\mathrm{visc}}$ and $f\equiv\mathrm{Re}(\omega)/2\pi$, as a function of the central bulk viscosity. Right: corresponding damping times in milliseconds, where $\tau\equiv-1/\mathrm{Im}(\omega)$. Causal frames A, B, and C are defined in Eq.~$(11)$ of Ref.~\cite{Keeble:2026bzo}, and, in this context, Eckart is equivalent to what we refer to as acausal Navier--Stokes. The underlying modes were computed using the frequency-domain approach of this work. This figure can be compared to Fig.~$3$ of Ref.~\cite{Keeble:2026bzo}, which was constructed from time-domain simulations.}
    \label{fig:frequency_shifts_zeta_extended}
\end{figure*}

\begin{table*}[hbt!]
\caption{\label{tbl:FDFreqs} Fundamental mode frequencies $f\equiv\mathrm{Re}(\omega)/2\pi$ (kHz) and damping times $\tau\equiv-1/\mathrm{Im}(\omega)$ (ms) for stars with central density
$\epsilon_{c}=5.5\times10^{15}\epscgs$ and viscosities 
$\zeta_{c} = 1.9\,\hat{\zeta}\times10^{31}\viscgs$, obtained from the tabulated time-domain modes in Ref.~\cite{Keeble:2026bzo} and frequency-domain computations using the approach described in this work. Here, causal frame A defined in Eq.~$(11)$ of Ref.~\cite{Keeble:2026bzo} is used. The
$(\cdot)$ notation denotes the first uncertain digit.}
\begin{ruledtabular}
\begin{tabular*}{\textwidth}{@{\extracolsep{\fill}}lcccc}
 & \multicolumn{2}{c}{\textrm{BDNK TD}} & \multicolumn{2}{c}{\textrm{BDNK FD}} \\
\cline{2-3} \cline{4-5}
 \multicolumn{1}{c}{$\hat{\zeta}$} & \multicolumn{1}{c}{$f$} &
 \multicolumn{1}{c}{$\tau$} & \multicolumn{1}{c}{$f$} &
 \multicolumn{1}{c}{$\tau$} \\ \hline
$0.01489375$ & $0.559135$ & $12.622(0)64$ & $0.558897(8)$ & $14.2130(8)$ \\
$0.02010656$ & $0.55905(0)$ & $9.511(4)23$ & $0.558806(4)$ & $10.4676(3)$ \\
$0.02714385$ & $0.55888(9)$ & $7.094(3)59$ & $0.558638(6)$ & $7.67313(1)$ \\
$0.0366442$ & $0.55857(8)$ & $5.22(6)969$ & $0.558328(6)$ & $5.577591(7)$ \\
\end{tabular*}
\end{ruledtabular}
\end{table*}

\appendix

\section{Derivation of the flat-space estimate}\label{app:flatspace}
For stars compact enough that $c_{s}^2\geq 1/2$, the second condition in Eq.~\eqref{eq:causalityb} implies $c_{s}^2\tau_{e}-\tau_{p}\geq 0$ and the right-hand side of Eq.~\eqref{eq:MinkFrameCorrection} is non-negative. From the second condition of Eq.~\eqref{eq:causalityb}, we have
\begin{align}
    \Delta\geq \frac{\nu+c_{s}^2\tau_{e}-\tau_{p}}{\left(1-c_{s}^2\right)\tau_{e}-\tau_p}\nu\tau_{e}k^2,
\end{align}
the right-hand side of which is an increasing function of $\tau_{p}$ if $c_{s}^2\geq 1/2$. The first condition of Eq.~\eqref{eq:causalityb} then gives
\begin{align}
    \Delta\geq \frac{c_{s}^2\tau_{e}^2}{\left(1-c_{s}^2\right)\tau_{e}-\nu}\nu k^2,
\end{align}
the right-hand side of which is minimized with respect to $\tau_{e}$ when $\tau_{e}=2\nu/(1-c_{s}^2)$. Hence, for $1/2\leq c_{s}^2<1$,
\begin{align}
    \Delta\geq \frac{4c_{s}^2\nu^{2}k^2}{\left(1-c_{s}^2\right)^2}.
\end{align}
It follows that the minimum frame deviation $\Delta_{\min}$ scales as $\zeta^{2}$. In particular,
\begin{align}
     \zeta^{*}\sim\sqrt{\Delta_{\min}}\frac{1-c_{s}^2}{2c_{s}}\frac{\epsilon+p}{k}.
\end{align}

\section{Consistency with time-domain code}\label{app:TD}

Figure~\ref{fig:frequency_shifts_zeta_extended} is an extension of Fig.~$3$ of Ref.~\cite{Keeble:2026bzo} to larger viscosities that were inaccessible in the time-domain simulations presented therein due to numerical instabilities. We find good agreement between the two approaches, with the frequency-domain results curing the numerical-viscosity-driven plateau at small viscosities. We show in Table~\ref{tbl:FDFreqs} a comparison with the tabulated time-domain modes presented in Ref.~\cite{Keeble:2026bzo}, also showing good agreement in the oscillation frequency. The time-domain modes overestimate the damping rate due to numerical viscosity.

\clearpage
\bibliographystyle{apsrev4-1}
\bibliography{refs}

\begin{thebibliography}{90}%
\makeatletter
\providecommand \@ifxundefined [1]{%
 \@ifx{#1\undefined}
}%
\providecommand \@ifnum [1]{%
 \ifnum #1\expandafter \@firstoftwo
 \else \expandafter \@secondoftwo
 \fi
}%
\providecommand \@ifx [1]{%
 \ifx #1\expandafter \@firstoftwo
 \else \expandafter \@secondoftwo
 \fi
}%
\providecommand \natexlab [1]{#1}%
\providecommand \enquote  [1]{``#1''}%
\providecommand \bibnamefont  [1]{#1}%
\providecommand \bibfnamefont [1]{#1}%
\providecommand \citenamefont [1]{#1}%
\providecommand \href@noop [0]{\@secondoftwo}%
\providecommand \href [0]{\begingroup \@sanitize@url \@href}%
\providecommand \@href[1]{\@@startlink{#1}\@@href}%
\providecommand \@@href[1]{\endgroup#1\@@endlink}%
\providecommand \@sanitize@url [0]{\catcode `\\12\catcode `\$12\catcode
  `\&12\catcode `\#12\catcode `\^12\catcode `\_12\catcode `\%12\relax}%
\providecommand \@@startlink[1]{}%
\providecommand \@@endlink[0]{}%
\providecommand \url  [0]{\begingroup\@sanitize@url \@url }%
\providecommand \@url [1]{\endgroup\@href {#1}{\urlprefix }}%
\providecommand \urlprefix  [0]{URL }%
\providecommand \Eprint [0]{\href }%
\providecommand \doibase [0]{http://dx.doi.org/}%
\providecommand \selectlanguage [0]{\@gobble}%
\providecommand \bibinfo  [0]{\@secondoftwo}%
\providecommand \bibfield  [0]{\@secondoftwo}%
\providecommand \translation [1]{[#1]}%
\providecommand \BibitemOpen [0]{}%
\providecommand \bibitemStop [0]{}%
\providecommand \bibitemNoStop [0]{.\EOS\space}%
\providecommand \EOS [0]{\spacefactor3000\relax}%
\providecommand \BibitemShut  [1]{\csname bibitem#1\endcsname}%
\let\auto@bib@innerbib\@empty
\bibitem [{\citenamefont {Andersson}\ and\ \citenamefont
  {Kokkotas}(1998)}]{Andersson:1997rn}%
  \BibitemOpen
  \bibfield  {author} {\bibinfo {author} {\bibfnamefont {N.}~\bibnamefont
  {Andersson}}\ and\ \bibinfo {author} {\bibfnamefont {K.~D.}\ \bibnamefont
  {Kokkotas}},\ }\href {\doibase 10.1046/j.1365-8711.1998.01840.x} {\bibfield
  {journal} {\bibinfo  {journal} {Mon. Not. Roy. Astron. Soc.}\ }\textbf
  {\bibinfo {volume} {299}},\ \bibinfo {pages} {1059} (\bibinfo {year}
  {1998})},\ \Eprint {http://arxiv.org/abs/gr-qc/9711088} {arXiv:gr-qc/9711088}
  \BibitemShut {NoStop}%
\bibitem [{\citenamefont {Kokkotas}\ and\ \citenamefont
  {Schmidt}(1999)}]{Kokkotas:1999bd}%
  \BibitemOpen
  \bibfield  {author} {\bibinfo {author} {\bibfnamefont {K.~D.}\ \bibnamefont
  {Kokkotas}}\ and\ \bibinfo {author} {\bibfnamefont {B.~G.}\ \bibnamefont
  {Schmidt}},\ }\href {\doibase 10.12942/lrr-1999-2} {\bibfield  {journal}
  {\bibinfo  {journal} {Living Rev. Rel.}\ }\textbf {\bibinfo {volume} {2}},\
  \bibinfo {pages} {2} (\bibinfo {year} {1999})},\ \Eprint
  {http://arxiv.org/abs/gr-qc/9909058} {arXiv:gr-qc/9909058} \BibitemShut
  {NoStop}%
\bibitem [{\citenamefont {Benhar}\ \emph {et~al.}(2004)\citenamefont {Benhar},
  \citenamefont {Ferrari},\ and\ \citenamefont {Gualtieri}}]{Benhar:2004xg}%
  \BibitemOpen
  \bibfield  {author} {\bibinfo {author} {\bibfnamefont {O.}~\bibnamefont
  {Benhar}}, \bibinfo {author} {\bibfnamefont {V.}~\bibnamefont {Ferrari}}, \
  and\ \bibinfo {author} {\bibfnamefont {L.}~\bibnamefont {Gualtieri}},\ }\href
  {\doibase 10.1103/PhysRevD.70.124015} {\bibfield  {journal} {\bibinfo
  {journal} {Phys. Rev. D}\ }\textbf {\bibinfo {volume} {70}},\ \bibinfo
  {pages} {124015} (\bibinfo {year} {2004})},\ \Eprint
  {http://arxiv.org/abs/astro-ph/0407529} {arXiv:astro-ph/0407529} \BibitemShut
  {NoStop}%
\bibitem [{\citenamefont {Lai}(1994)}]{Lai:1993di}%
  \BibitemOpen
  \bibfield  {author} {\bibinfo {author} {\bibfnamefont {D.}~\bibnamefont
  {Lai}},\ }\href {\doibase 10.1093/mnras/270.3.611} {\bibfield  {journal}
  {\bibinfo  {journal} {Mon. Not. Roy. Astron. Soc.}\ }\textbf {\bibinfo
  {volume} {270}},\ \bibinfo {pages} {611} (\bibinfo {year} {1994})},\ \Eprint
  {http://arxiv.org/abs/astro-ph/9404062} {arXiv:astro-ph/9404062} \BibitemShut
  {NoStop}%
\bibitem [{\citenamefont {Kokkotas}\ and\ \citenamefont
  {Schaefer}(1995)}]{Kokkotas:1995xe}%
  \BibitemOpen
  \bibfield  {author} {\bibinfo {author} {\bibfnamefont {K.~D.}\ \bibnamefont
  {Kokkotas}}\ and\ \bibinfo {author} {\bibfnamefont {G.}~\bibnamefont
  {Schaefer}},\ }\href {\doibase 10.1093/mnras/275.2.301} {\bibfield  {journal}
  {\bibinfo  {journal} {Mon. Not. Roy. Astron. Soc.}\ }\textbf {\bibinfo
  {volume} {275}},\ \bibinfo {pages} {301} (\bibinfo {year} {1995})},\ \Eprint
  {http://arxiv.org/abs/gr-qc/9502034} {arXiv:gr-qc/9502034} \BibitemShut
  {NoStop}%
\bibitem [{\citenamefont {Ho}\ and\ \citenamefont {Lai}(1999)}]{Ho:1998hq}%
  \BibitemOpen
  \bibfield  {author} {\bibinfo {author} {\bibfnamefont {W.~C.~G.}\
  \bibnamefont {Ho}}\ and\ \bibinfo {author} {\bibfnamefont {D.}~\bibnamefont
  {Lai}},\ }\href {\doibase 10.1046/j.1365-8711.1999.02703.x} {\bibfield
  {journal} {\bibinfo  {journal} {Mon. Not. Roy. Astron. Soc.}\ }\textbf
  {\bibinfo {volume} {308}},\ \bibinfo {pages} {153} (\bibinfo {year}
  {1999})},\ \Eprint {http://arxiv.org/abs/astro-ph/9812116}
  {arXiv:astro-ph/9812116} \BibitemShut {NoStop}%
\bibitem [{\citenamefont {Hinderer}\ \emph {et~al.}(2016)\citenamefont
  {Hinderer} \emph {et~al.}}]{Hinderer:2016eia}%
  \BibitemOpen
  \bibfield  {author} {\bibinfo {author} {\bibfnamefont {T.}~\bibnamefont
  {Hinderer}} \emph {et~al.},\ }\href {\doibase 10.1103/PhysRevLett.116.181101}
  {\bibfield  {journal} {\bibinfo  {journal} {Phys. Rev. Lett.}\ }\textbf
  {\bibinfo {volume} {116}},\ \bibinfo {pages} {181101} (\bibinfo {year}
  {2016})},\ \Eprint {http://arxiv.org/abs/1602.00599} {arXiv:1602.00599
  [gr-qc]} \BibitemShut {NoStop}%
\bibitem [{\citenamefont {Steinhoff}\ \emph {et~al.}(2016)\citenamefont
  {Steinhoff}, \citenamefont {Hinderer}, \citenamefont {Buonanno},\ and\
  \citenamefont {Taracchini}}]{Steinhoff:2016rfi}%
  \BibitemOpen
  \bibfield  {author} {\bibinfo {author} {\bibfnamefont {J.}~\bibnamefont
  {Steinhoff}}, \bibinfo {author} {\bibfnamefont {T.}~\bibnamefont {Hinderer}},
  \bibinfo {author} {\bibfnamefont {A.}~\bibnamefont {Buonanno}}, \ and\
  \bibinfo {author} {\bibfnamefont {A.}~\bibnamefont {Taracchini}},\ }\href
  {\doibase 10.1103/PhysRevD.94.104028} {\bibfield  {journal} {\bibinfo
  {journal} {Phys. Rev. D}\ }\textbf {\bibinfo {volume} {94}},\ \bibinfo
  {pages} {104028} (\bibinfo {year} {2016})},\ \Eprint
  {http://arxiv.org/abs/1608.01907} {arXiv:1608.01907 [gr-qc]} \BibitemShut
  {NoStop}%
\bibitem [{\citenamefont {Pratten}\ \emph {et~al.}(2020)\citenamefont
  {Pratten}, \citenamefont {Schmidt},\ and\ \citenamefont
  {Hinderer}}]{Pratten:2019sed}%
  \BibitemOpen
  \bibfield  {author} {\bibinfo {author} {\bibfnamefont {G.}~\bibnamefont
  {Pratten}}, \bibinfo {author} {\bibfnamefont {P.}~\bibnamefont {Schmidt}}, \
  and\ \bibinfo {author} {\bibfnamefont {T.}~\bibnamefont {Hinderer}},\ }\href
  {\doibase 10.1038/s41467-020-15984-5} {\bibfield  {journal} {\bibinfo
  {journal} {Nature Commun.}\ }\textbf {\bibinfo {volume} {11}},\ \bibinfo
  {pages} {2553} (\bibinfo {year} {2020})},\ \Eprint
  {http://arxiv.org/abs/1905.00817} {arXiv:1905.00817 [gr-qc]} \BibitemShut
  {NoStop}%
\bibitem [{\citenamefont {Alford}\ \emph {et~al.}(2018)\citenamefont {Alford},
  \citenamefont {Bovard}, \citenamefont {Hanauske}, \citenamefont {Rezzolla},\
  and\ \citenamefont {Schwenzer}}]{Alford:2017rxf}%
  \BibitemOpen
  \bibfield  {author} {\bibinfo {author} {\bibfnamefont {M.~G.}\ \bibnamefont
  {Alford}}, \bibinfo {author} {\bibfnamefont {L.}~\bibnamefont {Bovard}},
  \bibinfo {author} {\bibfnamefont {M.}~\bibnamefont {Hanauske}}, \bibinfo
  {author} {\bibfnamefont {L.}~\bibnamefont {Rezzolla}}, \ and\ \bibinfo
  {author} {\bibfnamefont {K.}~\bibnamefont {Schwenzer}},\ }\href {\doibase
  10.1103/PhysRevLett.120.041101} {\bibfield  {journal} {\bibinfo  {journal}
  {Phys. Rev. Lett.}\ }\textbf {\bibinfo {volume} {120}},\ \bibinfo {pages}
  {041101} (\bibinfo {year} {2018})},\ \Eprint
  {http://arxiv.org/abs/1707.09475} {arXiv:1707.09475 [gr-qc]} \BibitemShut
  {NoStop}%
\bibitem [{\citenamefont {Alford}\ \emph {et~al.}(2020)\citenamefont {Alford},
  \citenamefont {Harutyunyan},\ and\ \citenamefont
  {Sedrakian}}]{Alford:2020lla}%
  \BibitemOpen
  \bibfield  {author} {\bibinfo {author} {\bibfnamefont {M.}~\bibnamefont
  {Alford}}, \bibinfo {author} {\bibfnamefont {A.}~\bibnamefont {Harutyunyan}},
  \ and\ \bibinfo {author} {\bibfnamefont {A.}~\bibnamefont {Sedrakian}},\
  }\href {\doibase 10.3390/particles3020034} {\bibfield  {journal} {\bibinfo
  {journal} {Particles}\ }\textbf {\bibinfo {volume} {3}},\ \bibinfo {pages}
  {500} (\bibinfo {year} {2020})},\ \Eprint {http://arxiv.org/abs/2006.07975}
  {arXiv:2006.07975 [nucl-th]} \BibitemShut {NoStop}%
\bibitem [{\citenamefont {Most}\ \emph {et~al.}(2021)\citenamefont {Most},
  \citenamefont {Harris}, \citenamefont {Plumberg}, \citenamefont {Alford},
  \citenamefont {Noronha}, \citenamefont {Noronha-Hostler}, \citenamefont
  {Pretorius}, \citenamefont {Witek},\ and\ \citenamefont
  {Yunes}}]{Most:2021zvc}%
  \BibitemOpen
  \bibfield  {author} {\bibinfo {author} {\bibfnamefont {E.~R.}\ \bibnamefont
  {Most}}, \bibinfo {author} {\bibfnamefont {S.~P.}\ \bibnamefont {Harris}},
  \bibinfo {author} {\bibfnamefont {C.}~\bibnamefont {Plumberg}}, \bibinfo
  {author} {\bibfnamefont {M.~G.}\ \bibnamefont {Alford}}, \bibinfo {author}
  {\bibfnamefont {J.}~\bibnamefont {Noronha}}, \bibinfo {author} {\bibfnamefont
  {J.}~\bibnamefont {Noronha-Hostler}}, \bibinfo {author} {\bibfnamefont
  {F.}~\bibnamefont {Pretorius}}, \bibinfo {author} {\bibfnamefont
  {H.}~\bibnamefont {Witek}}, \ and\ \bibinfo {author} {\bibfnamefont
  {N.}~\bibnamefont {Yunes}},\ }\href {\doibase 10.1093/mnras/stab2793}
  {\bibfield  {journal} {\bibinfo  {journal} {Mon. Not. Roy. Astron. Soc.}\
  }\textbf {\bibinfo {volume} {509}},\ \bibinfo {pages} {1096} (\bibinfo {year}
  {2021})},\ \Eprint {http://arxiv.org/abs/2107.05094} {arXiv:2107.05094
  [astro-ph.HE]} \BibitemShut {NoStop}%
\bibitem [{\citenamefont {Most}\ \emph {et~al.}(2024)\citenamefont {Most},
  \citenamefont {Haber}, \citenamefont {Harris}, \citenamefont {Zhang},
  \citenamefont {Alford},\ and\ \citenamefont {Noronha}}]{Most:2022yhe}%
  \BibitemOpen
  \bibfield  {author} {\bibinfo {author} {\bibfnamefont {E.~R.}\ \bibnamefont
  {Most}}, \bibinfo {author} {\bibfnamefont {A.}~\bibnamefont {Haber}},
  \bibinfo {author} {\bibfnamefont {S.~P.}\ \bibnamefont {Harris}}, \bibinfo
  {author} {\bibfnamefont {Z.}~\bibnamefont {Zhang}}, \bibinfo {author}
  {\bibfnamefont {M.~G.}\ \bibnamefont {Alford}}, \ and\ \bibinfo {author}
  {\bibfnamefont {J.}~\bibnamefont {Noronha}},\ }\href {\doibase
  10.3847/2041-8213/ad454f} {\bibfield  {journal} {\bibinfo  {journal}
  {Astrophys. J. Lett.}\ }\textbf {\bibinfo {volume} {967}},\ \bibinfo {pages}
  {L14} (\bibinfo {year} {2024})},\ \Eprint {http://arxiv.org/abs/2207.00442}
  {arXiv:2207.00442 [astro-ph.HE]} \BibitemShut {NoStop}%
\bibitem [{\citenamefont {Ghosh}\ \emph {et~al.}(2024)\citenamefont {Ghosh},
  \citenamefont {Pradhan},\ and\ \citenamefont {Chatterjee}}]{Ghosh:2023vrx}%
  \BibitemOpen
  \bibfield  {author} {\bibinfo {author} {\bibfnamefont {S.}~\bibnamefont
  {Ghosh}}, \bibinfo {author} {\bibfnamefont {B.~K.}\ \bibnamefont {Pradhan}},
  \ and\ \bibinfo {author} {\bibfnamefont {D.}~\bibnamefont {Chatterjee}},\
  }\href {\doibase 10.1103/PhysRevD.109.103036} {\bibfield  {journal} {\bibinfo
   {journal} {Phys. Rev. D}\ }\textbf {\bibinfo {volume} {109}},\ \bibinfo
  {pages} {103036} (\bibinfo {year} {2024})},\ \Eprint
  {http://arxiv.org/abs/2306.14737} {arXiv:2306.14737 [gr-qc]} \BibitemShut
  {NoStop}%
\bibitem [{\citenamefont {Ghosh}\ \emph
  {et~al.}(2025{\natexlab{a}})\citenamefont {Ghosh}, \citenamefont {Mukherjee},
  \citenamefont {Bose},\ and\ \citenamefont {Chatterjee}}]{Ghosh:2025glz}%
  \BibitemOpen
  \bibfield  {author} {\bibinfo {author} {\bibfnamefont {S.}~\bibnamefont
  {Ghosh}}, \bibinfo {author} {\bibfnamefont {S.}~\bibnamefont {Mukherjee}},
  \bibinfo {author} {\bibfnamefont {S.}~\bibnamefont {Bose}}, \ and\ \bibinfo
  {author} {\bibfnamefont {D.}~\bibnamefont {Chatterjee}},\ }\href {\doibase
  10.1093/mnras/staf1652} {\bibfield  {journal} {\bibinfo  {journal} {Mon. Not.
  Roy. Astron. Soc.}\ }\textbf {\bibinfo {volume} {543}},\ \bibinfo {pages}
  {2987} (\bibinfo {year} {2025}{\natexlab{a}})},\ \Eprint
  {http://arxiv.org/abs/2503.14606} {arXiv:2503.14606 [gr-qc]} \BibitemShut
  {NoStop}%
\bibitem [{\citenamefont {Ghosh}\ \emph
  {et~al.}(2025{\natexlab{b}})\citenamefont {Ghosh}, \citenamefont
  {Hern{\'a}ndez}, \citenamefont {Pradhan}, \citenamefont {Manuel},
  \citenamefont {Chatterjee},\ and\ \citenamefont {Tolos}}]{Ghosh:2025wfx}%
  \BibitemOpen
  \bibfield  {author} {\bibinfo {author} {\bibfnamefont {S.}~\bibnamefont
  {Ghosh}}, \bibinfo {author} {\bibfnamefont {J.~L.}\ \bibnamefont
  {Hern{\'a}ndez}}, \bibinfo {author} {\bibfnamefont {B.~K.}\ \bibnamefont
  {Pradhan}}, \bibinfo {author} {\bibfnamefont {C.}~\bibnamefont {Manuel}},
  \bibinfo {author} {\bibfnamefont {D.}~\bibnamefont {Chatterjee}}, \ and\
  \bibinfo {author} {\bibfnamefont {L.}~\bibnamefont {Tolos}},\ }\href
  {\doibase 10.1103/h9r7-ld41} {\bibfield  {journal} {\bibinfo  {journal}
  {Phys. Rev. D}\ }\textbf {\bibinfo {volume} {112}},\ \bibinfo {pages}
  {084072} (\bibinfo {year} {2025}{\natexlab{b}})},\ \Eprint
  {http://arxiv.org/abs/2504.07659} {arXiv:2504.07659 [gr-qc]} \BibitemShut
  {NoStop}%
\bibitem [{\citenamefont {{Jones}}(2001)}]{2001PhRvD..64h4003J}%
  \BibitemOpen
  \bibfield  {author} {\bibinfo {author} {\bibfnamefont {P.~B.}\ \bibnamefont
  {{Jones}}},\ }\href {\doibase 10.1103/PhysRevD.64.084003} {\bibfield
  {journal} {\bibinfo  {journal} {\prd}\ }\textbf {\bibinfo {volume} {64}},\
  \bibinfo {eid} {084003} (\bibinfo {year} {2001})}\BibitemShut {NoStop}%
\bibitem [{\citenamefont {Alford}\ and\ \citenamefont
  {Haber}(2021)}]{Alford:2020pld}%
  \BibitemOpen
  \bibfield  {author} {\bibinfo {author} {\bibfnamefont {M.~G.}\ \bibnamefont
  {Alford}}\ and\ \bibinfo {author} {\bibfnamefont {A.}~\bibnamefont {Haber}},\
  }\href {\doibase 10.1103/PhysRevC.103.045810} {\bibfield  {journal} {\bibinfo
   {journal} {Phys. Rev. C}\ }\textbf {\bibinfo {volume} {103}},\ \bibinfo
  {pages} {045810} (\bibinfo {year} {2021})},\ \Eprint
  {http://arxiv.org/abs/2009.05181} {arXiv:2009.05181 [nucl-th]} \BibitemShut
  {NoStop}%
\bibitem [{\citenamefont {Gusakov}\ and\ \citenamefont
  {Kantor}(2008)}]{Gusakov:2008hv}%
  \BibitemOpen
  \bibfield  {author} {\bibinfo {author} {\bibfnamefont {M.~E.}\ \bibnamefont
  {Gusakov}}\ and\ \bibinfo {author} {\bibfnamefont {E.~M.}\ \bibnamefont
  {Kantor}},\ }\href {\doibase 10.1103/PhysRevD.78.083006} {\bibfield
  {journal} {\bibinfo  {journal} {Phys. Rev. D}\ }\textbf {\bibinfo {volume}
  {78}},\ \bibinfo {pages} {083006} (\bibinfo {year} {2008})},\ \Eprint
  {http://arxiv.org/abs/0806.4914} {arXiv:0806.4914 [astro-ph]} \BibitemShut
  {NoStop}%
\bibitem [{\citenamefont {Ripley}\ \emph {et~al.}(2024)\citenamefont {Ripley},
  \citenamefont {Hegade K.~R.}, \citenamefont {Chandramouli},\ and\
  \citenamefont {Yunes}}]{Ripley:2023lsq}%
  \BibitemOpen
  \bibfield  {author} {\bibinfo {author} {\bibfnamefont {J.~L.}\ \bibnamefont
  {Ripley}}, \bibinfo {author} {\bibfnamefont {A.}~\bibnamefont {Hegade
  K.~R.}}, \bibinfo {author} {\bibfnamefont {R.~S.}\ \bibnamefont
  {Chandramouli}}, \ and\ \bibinfo {author} {\bibfnamefont {N.}~\bibnamefont
  {Yunes}},\ }\href {\doibase 10.1038/s41550-024-02323-7} {\bibfield  {journal}
  {\bibinfo  {journal} {Nature Astron.}\ }\textbf {\bibinfo {volume} {8}},\
  \bibinfo {pages} {1277} (\bibinfo {year} {2024})},\ \Eprint
  {http://arxiv.org/abs/2312.11659} {arXiv:2312.11659 [gr-qc]} \BibitemShut
  {NoStop}%
\bibitem [{\citenamefont {Bildsten}\ and\ \citenamefont
  {Cutler}(1992)}]{Bildsten:1992my}%
  \BibitemOpen
  \bibfield  {author} {\bibinfo {author} {\bibfnamefont {L.}~\bibnamefont
  {Bildsten}}\ and\ \bibinfo {author} {\bibfnamefont {C.}~\bibnamefont
  {Cutler}},\ }\href {\doibase 10.1086/171983} {\bibfield  {journal} {\bibinfo
  {journal} {Astrophys. J.}\ }\textbf {\bibinfo {volume} {400}},\ \bibinfo
  {pages} {175} (\bibinfo {year} {1992})}\BibitemShut {NoStop}%
\bibitem [{\citenamefont {Kochanek}(1992)}]{Kochanek:1992wk}%
  \BibitemOpen
  \bibfield  {author} {\bibinfo {author} {\bibfnamefont {C.~S.}\ \bibnamefont
  {Kochanek}},\ }\href {\doibase 10.1086/171851} {\bibfield  {journal}
  {\bibinfo  {journal} {Astrophys. J.}\ }\textbf {\bibinfo {volume} {398}},\
  \bibinfo {pages} {234} (\bibinfo {year} {1992})}\BibitemShut {NoStop}%
\bibitem [{\citenamefont {Arras}\ and\ \citenamefont
  {Weinberg}(2019)}]{Arras:2018fxj}%
  \BibitemOpen
  \bibfield  {author} {\bibinfo {author} {\bibfnamefont {P.}~\bibnamefont
  {Arras}}\ and\ \bibinfo {author} {\bibfnamefont {N.~N.}\ \bibnamefont
  {Weinberg}},\ }\href {\doibase 10.1093/mnras/stz880} {\bibfield  {journal}
  {\bibinfo  {journal} {Mon. Not. Roy. Astron. Soc.}\ }\textbf {\bibinfo
  {volume} {486}},\ \bibinfo {pages} {1424} (\bibinfo {year} {2019})},\ \Eprint
  {http://arxiv.org/abs/1806.04163} {arXiv:1806.04163 [astro-ph.HE]}
  \BibitemShut {NoStop}%
\bibitem [{\citenamefont {{Lindblom}}\ and\ \citenamefont
  {{Hiscock}}(1983)}]{Lindblom_1983}%
  \BibitemOpen
  \bibfield  {author} {\bibinfo {author} {\bibfnamefont {L.}~\bibnamefont
  {{Lindblom}}}\ and\ \bibinfo {author} {\bibfnamefont {W.~A.}\ \bibnamefont
  {{Hiscock}}},\ }\href {\doibase 10.1086/160877} {\bibfield  {journal}
  {\bibinfo  {journal} {\apj}\ }\textbf {\bibinfo {volume} {267}},\ \bibinfo
  {pages} {384} (\bibinfo {year} {1983})}\BibitemShut {NoStop}%
\bibitem [{\citenamefont {Cutler}\ and\ \citenamefont
  {Lindblom}(1987)}]{Cutler:1987}%
  \BibitemOpen
  \bibfield  {author} {\bibinfo {author} {\bibfnamefont {C.}~\bibnamefont
  {Cutler}}\ and\ \bibinfo {author} {\bibfnamefont {L.}~\bibnamefont
  {Lindblom}},\ }\href {\doibase 10.1086/165052} {\bibfield  {journal}
  {\bibinfo  {journal} {Astrophys. J.}\ }\textbf {\bibinfo {volume} {314}},\
  \bibinfo {pages} {234} (\bibinfo {year} {1987})}\BibitemShut {NoStop}%
\bibitem [{\citenamefont {Sawyer}(1989)}]{Sawyer:1989dp}%
  \BibitemOpen
  \bibfield  {author} {\bibinfo {author} {\bibfnamefont {R.~F.}\ \bibnamefont
  {Sawyer}},\ }\href {\doibase 10.1103/PhysRevD.39.3804} {\bibfield  {journal}
  {\bibinfo  {journal} {Phys. Rev. D}\ }\textbf {\bibinfo {volume} {39}},\
  \bibinfo {pages} {3804} (\bibinfo {year} {1989})}\BibitemShut {NoStop}%
\bibitem [{\citenamefont {Cutler}\ \emph {et~al.}(1990)\citenamefont {Cutler},
  \citenamefont {Lindblom},\ and\ \citenamefont {Splinter}}]{Cutler:1990}%
  \BibitemOpen
  \bibfield  {author} {\bibinfo {author} {\bibfnamefont {C.}~\bibnamefont
  {Cutler}}, \bibinfo {author} {\bibfnamefont {L.}~\bibnamefont {Lindblom}}, \
  and\ \bibinfo {author} {\bibfnamefont {R.~J.}\ \bibnamefont {Splinter}},\
  }\href {\doibase 10.1086/169370} {\bibfield  {journal} {\bibinfo  {journal}
  {Astrophys. J.}\ }\textbf {\bibinfo {volume} {363}},\ \bibinfo {pages} {603}
  (\bibinfo {year} {1990})}\BibitemShut {NoStop}%
\bibitem [{\citenamefont {Gusakov}\ \emph {et~al.}(2005)\citenamefont
  {Gusakov}, \citenamefont {Yakovlev},\ and\ \citenamefont
  {Gnedin}}]{Gusakov:2005dz}%
  \BibitemOpen
  \bibfield  {author} {\bibinfo {author} {\bibfnamefont {M.~E.}\ \bibnamefont
  {Gusakov}}, \bibinfo {author} {\bibfnamefont {D.~G.}\ \bibnamefont
  {Yakovlev}}, \ and\ \bibinfo {author} {\bibfnamefont {O.~Y.}\ \bibnamefont
  {Gnedin}},\ }\href {\doibase 10.1111/j.1365-2966.2005.09295.x} {\bibfield
  {journal} {\bibinfo  {journal} {Mon. Not. Roy. Astron. Soc.}\ }\textbf
  {\bibinfo {volume} {361}},\ \bibinfo {pages} {1415} (\bibinfo {year}
  {2005})},\ \Eprint {http://arxiv.org/abs/astro-ph/0502583}
  {arXiv:astro-ph/0502583} \BibitemShut {NoStop}%
\bibitem [{\citenamefont {Ripley}\ \emph {et~al.}(2023)\citenamefont {Ripley},
  \citenamefont {Hegade K.~R.},\ and\ \citenamefont {Yunes}}]{Ripley:2023qxo}%
  \BibitemOpen
  \bibfield  {author} {\bibinfo {author} {\bibfnamefont {J.~L.}\ \bibnamefont
  {Ripley}}, \bibinfo {author} {\bibfnamefont {A.}~\bibnamefont {Hegade
  K.~R.}}, \ and\ \bibinfo {author} {\bibfnamefont {N.}~\bibnamefont {Yunes}},\
  }\href {\doibase 10.1103/PhysRevD.108.103037} {\bibfield  {journal} {\bibinfo
   {journal} {Phys. Rev. D}\ }\textbf {\bibinfo {volume} {108}},\ \bibinfo
  {pages} {103037} (\bibinfo {year} {2023})},\ \Eprint
  {http://arxiv.org/abs/2306.15633} {arXiv:2306.15633 [gr-qc]} \BibitemShut
  {NoStop}%
\bibitem [{\citenamefont {Hammond}\ \emph {et~al.}(2023)\citenamefont
  {Hammond}, \citenamefont {Hawke},\ and\ \citenamefont
  {Andersson}}]{Hammond:2022uua}%
  \BibitemOpen
  \bibfield  {author} {\bibinfo {author} {\bibfnamefont {P.}~\bibnamefont
  {Hammond}}, \bibinfo {author} {\bibfnamefont {I.}~\bibnamefont {Hawke}}, \
  and\ \bibinfo {author} {\bibfnamefont {N.}~\bibnamefont {Andersson}},\ }\href
  {\doibase 10.1103/PhysRevD.107.043023} {\bibfield  {journal} {\bibinfo
  {journal} {Phys. Rev. D}\ }\textbf {\bibinfo {volume} {107}},\ \bibinfo
  {pages} {043023} (\bibinfo {year} {2023})},\ \Eprint
  {http://arxiv.org/abs/2205.11377} {arXiv:2205.11377 [astro-ph.HE]}
  \BibitemShut {NoStop}%
\bibitem [{\citenamefont {Maggiore}\ \emph {et~al.}(2020)\citenamefont
  {Maggiore} \emph {et~al.}}]{ET:2019dnz}%
  \BibitemOpen
  \bibfield  {author} {\bibinfo {author} {\bibfnamefont {M.}~\bibnamefont
  {Maggiore}} \emph {et~al.} (\bibinfo {collaboration} {ET}),\ }\href {\doibase
  10.1088/1475-7516/2020/03/050} {\bibfield  {journal} {\bibinfo  {journal}
  {JCAP}\ }\textbf {\bibinfo {volume} {03}},\ \bibinfo {pages} {050} (\bibinfo
  {year} {2020})},\ \Eprint {http://arxiv.org/abs/1912.02622} {arXiv:1912.02622
  [astro-ph.CO]} \BibitemShut {NoStop}%
\bibitem [{\citenamefont {Abac}\ \emph {et~al.}(2026)\citenamefont {Abac} \emph
  {et~al.}}]{ET:2025xjr}%
  \BibitemOpen
  \bibfield  {author} {\bibinfo {author} {\bibfnamefont {A.}~\bibnamefont
  {Abac}} \emph {et~al.} (\bibinfo {collaboration} {ET}),\ }\href {\doibase
  10.1088/1475-7516/2026/03/081} {\bibfield  {journal} {\bibinfo  {journal}
  {JCAP}\ }\textbf {\bibinfo {volume} {03}},\ \bibinfo {pages} {081} (\bibinfo
  {year} {2026})},\ \Eprint {http://arxiv.org/abs/2503.12263} {arXiv:2503.12263
  [gr-qc]} \BibitemShut {NoStop}%
\bibitem [{\citenamefont {Reitze}\ \emph {et~al.}(2019)\citenamefont {Reitze}
  \emph {et~al.}}]{Reitze:2019iox}%
  \BibitemOpen
  \bibfield  {author} {\bibinfo {author} {\bibfnamefont {D.}~\bibnamefont
  {Reitze}} \emph {et~al.},\ }\href@noop {} {\bibfield  {journal} {\bibinfo
  {journal} {Bull. Am. Astron. Soc.}\ }\textbf {\bibinfo {volume} {51}},\
  \bibinfo {pages} {035} (\bibinfo {year} {2019})},\ \Eprint
  {http://arxiv.org/abs/1907.04833} {arXiv:1907.04833 [astro-ph.IM]}
  \BibitemShut {NoStop}%
\bibitem [{\citenamefont {Saketh}\ \emph {et~al.}(2024)\citenamefont {Saketh},
  \citenamefont {Zhou}, \citenamefont {Ghosh}, \citenamefont {Steinhoff},\ and\
  \citenamefont {Chatterjee}}]{Saketh:2024juq}%
  \BibitemOpen
  \bibfield  {author} {\bibinfo {author} {\bibfnamefont {M.~V.~S.}\
  \bibnamefont {Saketh}}, \bibinfo {author} {\bibfnamefont {Z.}~\bibnamefont
  {Zhou}}, \bibinfo {author} {\bibfnamefont {S.}~\bibnamefont {Ghosh}},
  \bibinfo {author} {\bibfnamefont {J.}~\bibnamefont {Steinhoff}}, \ and\
  \bibinfo {author} {\bibfnamefont {D.}~\bibnamefont {Chatterjee}},\ }\href
  {\doibase 10.1103/PhysRevD.110.103001} {\bibfield  {journal} {\bibinfo
  {journal} {Phys. Rev. D}\ }\textbf {\bibinfo {volume} {110}},\ \bibinfo
  {pages} {103001} (\bibinfo {year} {2024})},\ \Eprint
  {http://arxiv.org/abs/2407.08327} {arXiv:2407.08327 [gr-qc]} \BibitemShut
  {NoStop}%
\bibitem [{\citenamefont {Hegade K.~R.}\ \emph {et~al.}(2024)\citenamefont
  {Hegade K.~R.}, \citenamefont {Ripley},\ and\ \citenamefont
  {Yunes}}]{HegadeKR:2024agt}%
  \BibitemOpen
  \bibfield  {author} {\bibinfo {author} {\bibfnamefont {A.}~\bibnamefont
  {Hegade K.~R.}}, \bibinfo {author} {\bibfnamefont {J.~L.}\ \bibnamefont
  {Ripley}}, \ and\ \bibinfo {author} {\bibfnamefont {N.}~\bibnamefont
  {Yunes}},\ }\href {\doibase 10.1103/PhysRevD.109.104064} {\bibfield
  {journal} {\bibinfo  {journal} {Phys. Rev. D}\ }\textbf {\bibinfo {volume}
  {109}},\ \bibinfo {pages} {104064} (\bibinfo {year} {2024})},\ \Eprint
  {http://arxiv.org/abs/2403.03254} {arXiv:2403.03254 [gr-qc]} \BibitemShut
  {NoStop}%
\bibitem [{\citenamefont {Ghosh}\ \emph {et~al.}(2026)\citenamefont {Ghosh},
  \citenamefont {Haber}, \citenamefont {Andersson},\ and\ \citenamefont
  {Counsell}}]{Ghosh:2026ldg}%
  \BibitemOpen
  \bibfield  {author} {\bibinfo {author} {\bibfnamefont {S.}~\bibnamefont
  {Ghosh}}, \bibinfo {author} {\bibfnamefont {A.}~\bibnamefont {Haber}},
  \bibinfo {author} {\bibfnamefont {N.}~\bibnamefont {Andersson}}, \ and\
  \bibinfo {author} {\bibfnamefont {A.~R.}\ \bibnamefont {Counsell}},\
  }\href@noop {} {\  (\bibinfo {year} {2026})},\ \Eprint
  {http://arxiv.org/abs/2608.07311} {arXiv:2608.07311 [gr-qc]} \BibitemShut
  {NoStop}%
\bibitem [{\citenamefont {Caballero}\ and\ \citenamefont
  {Yunes}(2025)}]{Caballero_2025}%
  \BibitemOpen
  \bibfield  {author} {\bibinfo {author} {\bibfnamefont {D.~A.}\ \bibnamefont
  {Caballero}}\ and\ \bibinfo {author} {\bibfnamefont {N.}~\bibnamefont
  {Yunes}},\ }\href {\doibase 10.1103/cl4s-n7nr} {\bibfield  {journal}
  {\bibinfo  {journal} {Phys. Rev. D}\ }\textbf {\bibinfo {volume} {112}},\
  \bibinfo {pages} {063050} (\bibinfo {year} {2025})},\ \Eprint
  {http://arxiv.org/abs/2506.09149} {arXiv:2506.09149 [gr-qc]} \BibitemShut
  {NoStop}%
\bibitem [{\citenamefont {Mendes}\ \emph {et~al.}(2026)\citenamefont {Mendes},
  \citenamefont {Guerrieri}, \citenamefont {Muniz}, \citenamefont {Rocha},\
  and\ \citenamefont {Denicol}}]{Mendes:2025oib}%
  \BibitemOpen
  \bibfield  {author} {\bibinfo {author} {\bibfnamefont {R.~F.~P.}\
  \bibnamefont {Mendes}}, \bibinfo {author} {\bibfnamefont {A.}~\bibnamefont
  {Guerrieri}}, \bibinfo {author} {\bibfnamefont {J.~V.~M.}\ \bibnamefont
  {Muniz}}, \bibinfo {author} {\bibfnamefont {G.~S.}\ \bibnamefont {Rocha}}, \
  and\ \bibinfo {author} {\bibfnamefont {G.~S.}\ \bibnamefont {Denicol}},\
  }\href {\doibase 10.1103/fs15-pj7m} {\bibfield  {journal} {\bibinfo
  {journal} {Phys. Rev. D}\ }\textbf {\bibinfo {volume} {113}},\ \bibinfo
  {pages} {024010} (\bibinfo {year} {2026})},\ \Eprint
  {http://arxiv.org/abs/2509.12330} {arXiv:2509.12330 [gr-qc]} \BibitemShut
  {NoStop}%
\bibitem [{\citenamefont {Shum}\ \emph {et~al.}(2026)\citenamefont {Shum},
  \citenamefont {Abalos}, \citenamefont {Bea}, \citenamefont {Bezares},
  \citenamefont {Figueras},\ and\ \citenamefont {Palenzuela}}]{Shum:2025jnl}%
  \BibitemOpen
  \bibfield  {author} {\bibinfo {author} {\bibfnamefont {H.~L.~H.}\
  \bibnamefont {Shum}}, \bibinfo {author} {\bibfnamefont {F.}~\bibnamefont
  {Abalos}}, \bibinfo {author} {\bibfnamefont {Y.}~\bibnamefont {Bea}},
  \bibinfo {author} {\bibfnamefont {M.}~\bibnamefont {Bezares}}, \bibinfo
  {author} {\bibfnamefont {P.}~\bibnamefont {Figueras}}, \ and\ \bibinfo
  {author} {\bibfnamefont {C.}~\bibnamefont {Palenzuela}},\ }\href {\doibase
  10.1103/mthy-4119} {\bibfield  {journal} {\bibinfo  {journal} {Phys. Rev. D}\
  }\textbf {\bibinfo {volume} {113}},\ \bibinfo {pages} {084029} (\bibinfo
  {year} {2026})},\ \Eprint {http://arxiv.org/abs/2509.15303} {arXiv:2509.15303
  [gr-qc]} \BibitemShut {NoStop}%
\bibitem [{\citenamefont {Guerrieri}\ \emph {et~al.}(2026)\citenamefont
  {Guerrieri}, \citenamefont {Rocha}, \citenamefont {Denicol},\ and\
  \citenamefont {Mendes}}]{Guerrieri:2026jbm}%
  \BibitemOpen
  \bibfield  {author} {\bibinfo {author} {\bibfnamefont {A.}~\bibnamefont
  {Guerrieri}}, \bibinfo {author} {\bibfnamefont {G.~S.}\ \bibnamefont
  {Rocha}}, \bibinfo {author} {\bibfnamefont {G.~S.}\ \bibnamefont {Denicol}},
  \ and\ \bibinfo {author} {\bibfnamefont {R.~F.~P.}\ \bibnamefont {Mendes}},\
  }\href@noop {} {\  (\bibinfo {year} {2026})},\ \Eprint
  {http://arxiv.org/abs/2606.06600} {arXiv:2606.06600 [gr-qc]} \BibitemShut
  {NoStop}%
\bibitem [{\citenamefont {Denicol}\ \emph {et~al.}(2026)\citenamefont
  {Denicol}, \citenamefont {Guerrieri}, \citenamefont {Muniz}, \citenamefont
  {Rocha},\ and\ \citenamefont {Mendes}}]{Denicol:2026qqk}%
  \BibitemOpen
  \bibfield  {author} {\bibinfo {author} {\bibfnamefont {G.~S.}\ \bibnamefont
  {Denicol}}, \bibinfo {author} {\bibfnamefont {A.}~\bibnamefont {Guerrieri}},
  \bibinfo {author} {\bibfnamefont {J.~V.~M.}\ \bibnamefont {Muniz}}, \bibinfo
  {author} {\bibfnamefont {G.~S.}\ \bibnamefont {Rocha}}, \ and\ \bibinfo
  {author} {\bibfnamefont {R.~F.~P.}\ \bibnamefont {Mendes}},\ }\href@noop {}
  {\  (\bibinfo {year} {2026})},\ \Eprint {http://arxiv.org/abs/2608.19334}
  {arXiv:2608.19334 [gr-qc]} \BibitemShut {NoStop}%
\bibitem [{\citenamefont {Bussi{\`e}res}\ \emph {et~al.}(2026)\citenamefont
  {Bussi{\`e}res}, \citenamefont {Redondo-Yuste}, \citenamefont
  {Ortega~G{\'o}mez},\ and\ \citenamefont {Cardoso}}]{Bussieres:2026rnz}%
  \BibitemOpen
  \bibfield  {author} {\bibinfo {author} {\bibfnamefont {S.}~\bibnamefont
  {Bussi{\`e}res}}, \bibinfo {author} {\bibfnamefont {J.}~\bibnamefont
  {Redondo-Yuste}}, \bibinfo {author} {\bibfnamefont {J.~J.}\ \bibnamefont
  {Ortega~G{\'o}mez}}, \ and\ \bibinfo {author} {\bibfnamefont
  {V.}~\bibnamefont {Cardoso}},\ }\href {\doibase 10.1103/w9nm-ygnf} {\bibfield
   {journal} {\bibinfo  {journal} {Phys. Rev. D}\ }\textbf {\bibinfo {volume}
  {114}},\ \bibinfo {pages} {044012} (\bibinfo {year} {2026})},\ \Eprint
  {http://arxiv.org/abs/2604.13208} {arXiv:2604.13208 [gr-qc]} \BibitemShut
  {NoStop}%
\bibitem [{\citenamefont {Katagiri}\ \emph {et~al.}(2026)\citenamefont
  {Katagiri}, \citenamefont {Guedes},\ and\ \citenamefont
  {Yagi}}]{Katagiri:2026jgp}%
  \BibitemOpen
  \bibfield  {author} {\bibinfo {author} {\bibfnamefont {T.}~\bibnamefont
  {Katagiri}}, \bibinfo {author} {\bibfnamefont {V.}~\bibnamefont {Guedes}}, \
  and\ \bibinfo {author} {\bibfnamefont {K.}~\bibnamefont {Yagi}},\ }\href@noop
  {} {\  (\bibinfo {year} {2026})},\ \Eprint {http://arxiv.org/abs/2606.20957}
  {arXiv:2606.20957 [gr-qc]} \BibitemShut {NoStop}%
\bibitem [{\citenamefont {Chabanov}\ and\ \citenamefont
  {Rezzolla}(2025{\natexlab{a}})}]{Chabanov:2023abq}%
  \BibitemOpen
  \bibfield  {author} {\bibinfo {author} {\bibfnamefont {M.}~\bibnamefont
  {Chabanov}}\ and\ \bibinfo {author} {\bibfnamefont {L.}~\bibnamefont
  {Rezzolla}},\ }\href {\doibase 10.1103/PhysRevD.111.044074} {\bibfield
  {journal} {\bibinfo  {journal} {Phys. Rev. D}\ }\textbf {\bibinfo {volume}
  {111}},\ \bibinfo {pages} {044074} (\bibinfo {year} {2025}{\natexlab{a}})},\
  \Eprint {http://arxiv.org/abs/2311.13027} {arXiv:2311.13027 [gr-qc]}
  \BibitemShut {NoStop}%
\bibitem [{\citenamefont {Chabanov}\ and\ \citenamefont
  {Rezzolla}(2025{\natexlab{b}})}]{Chabanov:2023blf}%
  \BibitemOpen
  \bibfield  {author} {\bibinfo {author} {\bibfnamefont {M.}~\bibnamefont
  {Chabanov}}\ and\ \bibinfo {author} {\bibfnamefont {L.}~\bibnamefont
  {Rezzolla}},\ }\href {\doibase 10.1103/PhysRevLett.134.071402} {\bibfield
  {journal} {\bibinfo  {journal} {Phys. Rev. Lett.}\ }\textbf {\bibinfo
  {volume} {134}},\ \bibinfo {pages} {071402} (\bibinfo {year}
  {2025}{\natexlab{b}})},\ \Eprint {http://arxiv.org/abs/2307.10464}
  {arXiv:2307.10464 [gr-qc]} \BibitemShut {NoStop}%
\bibitem [{\citenamefont {Boyanov}\ \emph {et~al.}(2025)\citenamefont
  {Boyanov}, \citenamefont {Cardoso}, \citenamefont {Kokkotas},\ and\
  \citenamefont {Redondo-Yuste}}]{Boyanov:2024}%
  \BibitemOpen
  \bibfield  {author} {\bibinfo {author} {\bibfnamefont {V.}~\bibnamefont
  {Boyanov}}, \bibinfo {author} {\bibfnamefont {V.}~\bibnamefont {Cardoso}},
  \bibinfo {author} {\bibfnamefont {K.~D.}\ \bibnamefont {Kokkotas}}, \ and\
  \bibinfo {author} {\bibfnamefont {J.}~\bibnamefont {Redondo-Yuste}},\ }\href
  {\doibase 10.1103/smlr-v7b2} {\bibfield  {journal} {\bibinfo  {journal}
  {Phys. Rev. Lett.}\ }\textbf {\bibinfo {volume} {135}},\ \bibinfo {pages}
  {151402} (\bibinfo {year} {2025})},\ \Eprint
  {http://arxiv.org/abs/2411.16861} {arXiv:2411.16861 [gr-qc]} \BibitemShut
  {NoStop}%
\bibitem [{\citenamefont {Redondo-Yuste}\ and\ \citenamefont
  {Cardoso}(2025)}]{Redondo-Yuste:2025ktt}%
  \BibitemOpen
  \bibfield  {author} {\bibinfo {author} {\bibfnamefont {J.}~\bibnamefont
  {Redondo-Yuste}}\ and\ \bibinfo {author} {\bibfnamefont {V.}~\bibnamefont
  {Cardoso}},\ }\href {\doibase 10.1103/gvd3-lqkv} {\bibfield  {journal}
  {\bibinfo  {journal} {Phys. Rev. D}\ }\textbf {\bibinfo {volume} {112}},\
  \bibinfo {pages} {L061501} (\bibinfo {year} {2025})},\ \Eprint
  {http://arxiv.org/abs/2506.13850} {arXiv:2506.13850 [gr-qc]} \BibitemShut
  {NoStop}%
\bibitem [{\citenamefont {Redondo-Yuste}(2025)}]{Redondo-Yuste:2024vdb}%
  \BibitemOpen
  \bibfield  {author} {\bibinfo {author} {\bibfnamefont {J.}~\bibnamefont
  {Redondo-Yuste}},\ }\href {\doibase 10.1088/1361-6382/adbfef} {\bibfield
  {journal} {\bibinfo  {journal} {Class. Quant. Grav.}\ }\textbf {\bibinfo
  {volume} {42}},\ \bibinfo {pages} {075012} (\bibinfo {year} {2025})},\
  \Eprint {http://arxiv.org/abs/2411.16841} {arXiv:2411.16841 [gr-qc]}
  \BibitemShut {NoStop}%
\bibitem [{\citenamefont {Keeble}\ and\ \citenamefont
  {Redondo-Yuste}(2026)}]{Keeble:2026bzo}%
  \BibitemOpen
  \bibfield  {author} {\bibinfo {author} {\bibfnamefont {L.~S.}\ \bibnamefont
  {Keeble}}\ and\ \bibinfo {author} {\bibfnamefont {J.}~\bibnamefont
  {Redondo-Yuste}},\ }\href {\doibase 10.1103/41kq-fj3h} {\bibfield  {journal}
  {\bibinfo  {journal} {Phys. Rev. D}\ }\textbf {\bibinfo {volume} {113}},\
  \bibinfo {pages} {124058} (\bibinfo {year} {2026})},\ \Eprint
  {http://arxiv.org/abs/2603.23622} {arXiv:2603.23622 [gr-qc]} \BibitemShut
  {NoStop}%
\bibitem [{\citenamefont {Camelio}\ \emph
  {et~al.}(2023{\natexlab{a}})\citenamefont {Camelio}, \citenamefont
  {Gavassino}, \citenamefont {Antonelli}, \citenamefont {Bernuzzi},\ and\
  \citenamefont {Haskell}}]{Camelio:2022ljs}%
  \BibitemOpen
  \bibfield  {author} {\bibinfo {author} {\bibfnamefont {G.}~\bibnamefont
  {Camelio}}, \bibinfo {author} {\bibfnamefont {L.}~\bibnamefont {Gavassino}},
  \bibinfo {author} {\bibfnamefont {M.}~\bibnamefont {Antonelli}}, \bibinfo
  {author} {\bibfnamefont {S.}~\bibnamefont {Bernuzzi}}, \ and\ \bibinfo
  {author} {\bibfnamefont {B.}~\bibnamefont {Haskell}},\ }\href {\doibase
  10.1103/PhysRevD.107.103031} {\bibfield  {journal} {\bibinfo  {journal}
  {Phys. Rev. D}\ }\textbf {\bibinfo {volume} {107}},\ \bibinfo {pages}
  {103031} (\bibinfo {year} {2023}{\natexlab{a}})},\ \Eprint
  {http://arxiv.org/abs/2204.11809} {arXiv:2204.11809 [gr-qc]} \BibitemShut
  {NoStop}%
\bibitem [{\citenamefont {Camelio}\ \emph
  {et~al.}(2023{\natexlab{b}})\citenamefont {Camelio}, \citenamefont
  {Gavassino}, \citenamefont {Antonelli}, \citenamefont {Bernuzzi},\ and\
  \citenamefont {Haskell}}]{Camelio:2022fds}%
  \BibitemOpen
  \bibfield  {author} {\bibinfo {author} {\bibfnamefont {G.}~\bibnamefont
  {Camelio}}, \bibinfo {author} {\bibfnamefont {L.}~\bibnamefont {Gavassino}},
  \bibinfo {author} {\bibfnamefont {M.}~\bibnamefont {Antonelli}}, \bibinfo
  {author} {\bibfnamefont {S.}~\bibnamefont {Bernuzzi}}, \ and\ \bibinfo
  {author} {\bibfnamefont {B.}~\bibnamefont {Haskell}},\ }\href {\doibase
  10.1103/PhysRevD.107.103032} {\bibfield  {journal} {\bibinfo  {journal}
  {Phys. Rev. D}\ }\textbf {\bibinfo {volume} {107}},\ \bibinfo {pages}
  {103032} (\bibinfo {year} {2023}{\natexlab{b}})},\ \Eprint
  {http://arxiv.org/abs/2204.11810} {arXiv:2204.11810 [gr-qc]} \BibitemShut
  {NoStop}%
\bibitem [{\citenamefont {Kovtun}(2012)}]{Kovtun:2012rj}%
  \BibitemOpen
  \bibfield  {author} {\bibinfo {author} {\bibfnamefont {P.}~\bibnamefont
  {Kovtun}},\ }\href {\doibase 10.1088/1751-8113/45/47/473001} {\bibfield
  {journal} {\bibinfo  {journal} {J. Phys. A}\ }\textbf {\bibinfo {volume}
  {45}},\ \bibinfo {pages} {473001} (\bibinfo {year} {2012})},\ \Eprint
  {http://arxiv.org/abs/1205.5040} {arXiv:1205.5040 [hep-th]} \BibitemShut
  {NoStop}%
\bibitem [{\citenamefont {Kovtun}(2019)}]{Kovtun:2019hdm}%
  \BibitemOpen
  \bibfield  {author} {\bibinfo {author} {\bibfnamefont {P.}~\bibnamefont
  {Kovtun}},\ }\href {\doibase 10.1007/JHEP10(2019)034} {\bibfield  {journal}
  {\bibinfo  {journal} {JHEP}\ }\textbf {\bibinfo {volume} {10}},\ \bibinfo
  {pages} {034} (\bibinfo {year} {2019})},\ \Eprint
  {http://arxiv.org/abs/1907.08191} {arXiv:1907.08191 [hep-th]} \BibitemShut
  {NoStop}%
\bibitem [{\citenamefont {Hoult}\ and\ \citenamefont
  {Kovtun}(2020)}]{Hoult:2020eho}%
  \BibitemOpen
  \bibfield  {author} {\bibinfo {author} {\bibfnamefont {R.~E.}\ \bibnamefont
  {Hoult}}\ and\ \bibinfo {author} {\bibfnamefont {P.}~\bibnamefont {Kovtun}},\
  }\href {\doibase 10.1007/JHEP06(2020)067} {\bibfield  {journal} {\bibinfo
  {journal} {JHEP}\ }\textbf {\bibinfo {volume} {06}},\ \bibinfo {pages} {067}
  (\bibinfo {year} {2020})},\ \Eprint {http://arxiv.org/abs/2004.04102}
  {arXiv:2004.04102 [hep-th]} \BibitemShut {NoStop}%
\bibitem [{\citenamefont {Rocha}\ \emph {et~al.}(2024)\citenamefont {Rocha},
  \citenamefont {Wagner}, \citenamefont {Denicol}, \citenamefont {Noronha},\
  and\ \citenamefont {Rischke}}]{Rocha:2023ilf}%
  \BibitemOpen
  \bibfield  {author} {\bibinfo {author} {\bibfnamefont {G.~S.}\ \bibnamefont
  {Rocha}}, \bibinfo {author} {\bibfnamefont {D.}~\bibnamefont {Wagner}},
  \bibinfo {author} {\bibfnamefont {G.~S.}\ \bibnamefont {Denicol}}, \bibinfo
  {author} {\bibfnamefont {J.}~\bibnamefont {Noronha}}, \ and\ \bibinfo
  {author} {\bibfnamefont {D.~H.}\ \bibnamefont {Rischke}},\ }\href {\doibase
  10.3390/e26030189} {\bibfield  {journal} {\bibinfo  {journal} {Entropy}\
  }\textbf {\bibinfo {volume} {26}},\ \bibinfo {pages} {189} (\bibinfo {year}
  {2024})},\ \Eprint {http://arxiv.org/abs/2311.15063} {arXiv:2311.15063
  [nucl-th]} \BibitemShut {NoStop}%
\bibitem [{\citenamefont {Disconzi}(2024)}]{disconzi_2024}%
  \BibitemOpen
  \bibfield  {author} {\bibinfo {author} {\bibfnamefont {M.~M.}\ \bibnamefont
  {Disconzi}},\ }\href {\doibase 10.1007/s41114-024-00052-x} {\bibfield
  {journal} {\bibinfo  {journal} {Living Rev. Rel.}\ }\textbf {\bibinfo
  {volume} {27}},\ \bibinfo {pages} {6} (\bibinfo {year} {2024})},\ \Eprint
  {http://arxiv.org/abs/2308.09844} {arXiv:2308.09844 [math.AP]} \BibitemShut
  {NoStop}%
\bibitem [{\citenamefont {Bemfica}\ \emph {et~al.}(2018)\citenamefont
  {Bemfica}, \citenamefont {Disconzi},\ and\ \citenamefont
  {Noronha}}]{Bemfica:2017wps}%
  \BibitemOpen
  \bibfield  {author} {\bibinfo {author} {\bibfnamefont {F.~S.}\ \bibnamefont
  {Bemfica}}, \bibinfo {author} {\bibfnamefont {M.~M.}\ \bibnamefont
  {Disconzi}}, \ and\ \bibinfo {author} {\bibfnamefont {J.}~\bibnamefont
  {Noronha}},\ }\href {\doibase 10.1103/PhysRevD.98.104064} {\bibfield
  {journal} {\bibinfo  {journal} {Phys. Rev. D}\ }\textbf {\bibinfo {volume}
  {98}},\ \bibinfo {pages} {104064} (\bibinfo {year} {2018})},\ \Eprint
  {http://arxiv.org/abs/1708.06255} {arXiv:1708.06255 [gr-qc]} \BibitemShut
  {NoStop}%
\bibitem [{\citenamefont {Bemfica}\ \emph
  {et~al.}(2019{\natexlab{a}})\citenamefont {Bemfica}, \citenamefont
  {Disconzi},\ and\ \citenamefont {Noronha}}]{Bemfica:2019knx}%
  \BibitemOpen
  \bibfield  {author} {\bibinfo {author} {\bibfnamefont {F.~S.}\ \bibnamefont
  {Bemfica}}, \bibinfo {author} {\bibfnamefont {M.~M.}\ \bibnamefont
  {Disconzi}}, \ and\ \bibinfo {author} {\bibfnamefont {J.}~\bibnamefont
  {Noronha}},\ }\href {\doibase 10.1103/PhysRevD.100.104020} {\bibfield
  {journal} {\bibinfo  {journal} {Phys. Rev. D}\ }\textbf {\bibinfo {volume}
  {100}},\ \bibinfo {pages} {104020} (\bibinfo {year} {2019}{\natexlab{a}})},\
  \bibinfo {note} {[Erratum: Phys. Rev. D 105, 069902 (2022)]},\ \Eprint
  {http://arxiv.org/abs/1907.12695} {arXiv:1907.12695 [gr-qc]} \BibitemShut
  {NoStop}%
\bibitem [{\citenamefont {Bemfica}\ \emph {et~al.}(2022)\citenamefont
  {Bemfica}, \citenamefont {Disconzi},\ and\ \citenamefont
  {Noronha}}]{Bemfica:2020zjp}%
  \BibitemOpen
  \bibfield  {author} {\bibinfo {author} {\bibfnamefont {F.~S.}\ \bibnamefont
  {Bemfica}}, \bibinfo {author} {\bibfnamefont {M.~M.}\ \bibnamefont
  {Disconzi}}, \ and\ \bibinfo {author} {\bibfnamefont {J.}~\bibnamefont
  {Noronha}},\ }\href {\doibase 10.1103/PhysRevX.12.021044} {\bibfield
  {journal} {\bibinfo  {journal} {Phys. Rev. X}\ }\textbf {\bibinfo {volume}
  {12}},\ \bibinfo {pages} {021044} (\bibinfo {year} {2022})},\ \Eprint
  {http://arxiv.org/abs/2009.11388} {arXiv:2009.11388 [gr-qc]} \BibitemShut
  {NoStop}%
\bibitem [{\citenamefont {{M{\"u}ller}}(1967)}]{Muller_1967}%
  \BibitemOpen
  \bibfield  {author} {\bibinfo {author} {\bibfnamefont {I.}~\bibnamefont
  {{M{\"u}ller}}},\ }\href {\doibase 10.1007/BF01326412} {\bibfield  {journal}
  {\bibinfo  {journal} {Z. Phys.}\ }\textbf {\bibinfo {volume} {198}},\
  \bibinfo {pages} {329} (\bibinfo {year} {1967})}\BibitemShut {NoStop}%
\bibitem [{\citenamefont {Israel}\ and\ \citenamefont
  {Stewart}(1976)}]{ISRAEL1976213}%
  \BibitemOpen
  \bibfield  {author} {\bibinfo {author} {\bibfnamefont {W.}~\bibnamefont
  {Israel}}\ and\ \bibinfo {author} {\bibfnamefont {J.~M.}\ \bibnamefont
  {Stewart}},\ }\href {\doibase 10.1016/0375-9601(76)90075-X} {\bibfield
  {journal} {\bibinfo  {journal} {Phys. Lett. A}\ }\textbf {\bibinfo {volume}
  {58}},\ \bibinfo {pages} {213} (\bibinfo {year} {1976})}\BibitemShut
  {NoStop}%
\bibitem [{\citenamefont {Israel}(1976)}]{ISRAEL1976310}%
  \BibitemOpen
  \bibfield  {author} {\bibinfo {author} {\bibfnamefont {W.}~\bibnamefont
  {Israel}},\ }\href {\doibase 10.1016/0003-4916(76)90064-6} {\bibfield
  {journal} {\bibinfo  {journal} {Annals Phys.}\ }\textbf {\bibinfo {volume}
  {100}},\ \bibinfo {pages} {310} (\bibinfo {year} {1976})}\BibitemShut
  {NoStop}%
\bibitem [{\citenamefont {Israel}\ and\ \citenamefont
  {Stewart}(1979)}]{ISRAEL1979341}%
  \BibitemOpen
  \bibfield  {author} {\bibinfo {author} {\bibfnamefont {W.}~\bibnamefont
  {Israel}}\ and\ \bibinfo {author} {\bibfnamefont {J.~M.}\ \bibnamefont
  {Stewart}},\ }\href {\doibase 10.1016/0003-4916(79)90130-1} {\bibfield
  {journal} {\bibinfo  {journal} {Annals Phys.}\ }\textbf {\bibinfo {volume}
  {118}},\ \bibinfo {pages} {341} (\bibinfo {year} {1979})}\BibitemShut
  {NoStop}%
\bibitem [{\citenamefont {Hiscock}\ and\ \citenamefont
  {Lindblom}(1983)}]{Hiscock:1983}%
  \BibitemOpen
  \bibfield  {author} {\bibinfo {author} {\bibfnamefont {W.~A.}\ \bibnamefont
  {Hiscock}}\ and\ \bibinfo {author} {\bibfnamefont {L.}~\bibnamefont
  {Lindblom}},\ }\href {\doibase 10.1016/0003-4916(83)90288-9} {\bibfield
  {journal} {\bibinfo  {journal} {Annals Phys.}\ }\textbf {\bibinfo {volume}
  {151}},\ \bibinfo {pages} {466} (\bibinfo {year} {1983})}\BibitemShut
  {NoStop}%
\bibitem [{\citenamefont {Olson}(1990)}]{Olson_1990}%
  \BibitemOpen
  \bibfield  {author} {\bibinfo {author} {\bibfnamefont {T.~S.}\ \bibnamefont
  {Olson}},\ }\href {\doibase 10.1016/0003-4916(90)90366-V} {\bibfield
  {journal} {\bibinfo  {journal} {Annals Phys.}\ }\textbf {\bibinfo {volume}
  {199}},\ \bibinfo {pages} {18} (\bibinfo {year} {1990})}\BibitemShut
  {NoStop}%
\bibitem [{\citenamefont {Cordeiro}\ \emph
  {et~al.}(2026{\natexlab{a}})\citenamefont {Cordeiro}, \citenamefont
  {Bemfica}, \citenamefont {Speranza},\ and\ \citenamefont
  {Noronha}}]{Cordeiro_2025}%
  \BibitemOpen
  \bibfield  {author} {\bibinfo {author} {\bibfnamefont {I.}~\bibnamefont
  {Cordeiro}}, \bibinfo {author} {\bibfnamefont {F.~S.}\ \bibnamefont
  {Bemfica}}, \bibinfo {author} {\bibfnamefont {E.}~\bibnamefont {Speranza}}, \
  and\ \bibinfo {author} {\bibfnamefont {J.}~\bibnamefont {Noronha}},\ }\href
  {\doibase 10.1103/dkcq-6s5b} {\bibfield  {journal} {\bibinfo  {journal}
  {Phys. Rev. D}\ }\textbf {\bibinfo {volume} {113}},\ \bibinfo {pages}
  {045024} (\bibinfo {year} {2026}{\natexlab{a}})},\ \Eprint
  {http://arxiv.org/abs/2507.20064} {arXiv:2507.20064 [nucl-th]} \BibitemShut
  {NoStop}%
\bibitem [{\citenamefont {Cordeiro}\ \emph
  {et~al.}(2026{\natexlab{b}})\citenamefont {Cordeiro}, \citenamefont
  {Bemfica}, \citenamefont {Speranza},\ and\ \citenamefont
  {Noronha}}]{Cordeiro:2025rkt}%
  \BibitemOpen
  \bibfield  {author} {\bibinfo {author} {\bibfnamefont {I.}~\bibnamefont
  {Cordeiro}}, \bibinfo {author} {\bibfnamefont {F.~S.}\ \bibnamefont
  {Bemfica}}, \bibinfo {author} {\bibfnamefont {E.}~\bibnamefont {Speranza}}, \
  and\ \bibinfo {author} {\bibfnamefont {J.}~\bibnamefont {Noronha}},\ }\href
  {\doibase 10.1051/epjconf/202636412003} {\bibfield  {journal} {\bibinfo
  {journal} {EPJ Web Conf.}\ }\textbf {\bibinfo {volume} {364}},\ \bibinfo
  {pages} {12003} (\bibinfo {year} {2026}{\natexlab{b}})},\ \Eprint
  {http://arxiv.org/abs/2510.01512} {arXiv:2510.01512 [nucl-th]} \BibitemShut
  {NoStop}%
\bibitem [{\citenamefont {Cordeiro}\ \emph
  {et~al.}(2026{\natexlab{c}})\citenamefont {Cordeiro}, \citenamefont
  {Speranza}, \citenamefont {Bemfica}, \citenamefont {Disconzi},\ and\
  \citenamefont {Noronha}}]{Cordeiro:2026jle}%
  \BibitemOpen
  \bibfield  {author} {\bibinfo {author} {\bibfnamefont {I.}~\bibnamefont
  {Cordeiro}}, \bibinfo {author} {\bibfnamefont {E.}~\bibnamefont {Speranza}},
  \bibinfo {author} {\bibfnamefont {F.~S.}\ \bibnamefont {Bemfica}}, \bibinfo
  {author} {\bibfnamefont {M.~M.}\ \bibnamefont {Disconzi}}, \ and\ \bibinfo
  {author} {\bibfnamefont {J.}~\bibnamefont {Noronha}},\ }\href@noop {} {\
  (\bibinfo {year} {2026}{\natexlab{c}})},\ \Eprint
  {http://arxiv.org/abs/2607.05639} {arXiv:2607.05639 [nucl-th]} \BibitemShut
  {NoStop}%
\bibitem [{\citenamefont {Bemfica}\ \emph
  {et~al.}(2019{\natexlab{b}})\citenamefont {Bemfica}, \citenamefont
  {Disconzi},\ and\ \citenamefont {Noronha}}]{Bemfica_2019b}%
  \BibitemOpen
  \bibfield  {author} {\bibinfo {author} {\bibfnamefont {F.~S.}\ \bibnamefont
  {Bemfica}}, \bibinfo {author} {\bibfnamefont {M.~M.}\ \bibnamefont
  {Disconzi}}, \ and\ \bibinfo {author} {\bibfnamefont {J.}~\bibnamefont
  {Noronha}},\ }\href {\doibase 10.1103/PhysRevLett.122.221602} {\bibfield
  {journal} {\bibinfo  {journal} {Phys. Rev. Lett.}\ }\textbf {\bibinfo
  {volume} {122}},\ \bibinfo {pages} {221602} (\bibinfo {year}
  {2019}{\natexlab{b}})},\ \Eprint {http://arxiv.org/abs/1901.06701}
  {arXiv:1901.06701 [gr-qc]} \BibitemShut {NoStop}%
\bibitem [{\citenamefont {Pu}\ \emph {et~al.}(2010)\citenamefont {Pu},
  \citenamefont {Koide},\ and\ \citenamefont {Rischke}}]{Pu:2009fj}%
  \BibitemOpen
  \bibfield  {author} {\bibinfo {author} {\bibfnamefont {S.}~\bibnamefont
  {Pu}}, \bibinfo {author} {\bibfnamefont {T.}~\bibnamefont {Koide}}, \ and\
  \bibinfo {author} {\bibfnamefont {D.~H.}\ \bibnamefont {Rischke}},\ }\href
  {\doibase 10.1103/PhysRevD.81.114039} {\bibfield  {journal} {\bibinfo
  {journal} {Phys. Rev. D}\ }\textbf {\bibinfo {volume} {81}},\ \bibinfo
  {pages} {114039} (\bibinfo {year} {2010})},\ \Eprint
  {http://arxiv.org/abs/0907.3906} {arXiv:0907.3906 [hep-ph]} \BibitemShut
  {NoStop}%
\bibitem [{\citenamefont {Denicol}\ \emph {et~al.}(2008)\citenamefont
  {Denicol}, \citenamefont {Kodama}, \citenamefont {Koide},\ and\ \citenamefont
  {Mota}}]{Denicol:2008ha}%
  \BibitemOpen
  \bibfield  {author} {\bibinfo {author} {\bibfnamefont {G.~S.}\ \bibnamefont
  {Denicol}}, \bibinfo {author} {\bibfnamefont {T.}~\bibnamefont {Kodama}},
  \bibinfo {author} {\bibfnamefont {T.}~\bibnamefont {Koide}}, \ and\ \bibinfo
  {author} {\bibfnamefont {P.}~\bibnamefont {Mota}},\ }\href {\doibase
  10.1088/0954-3899/35/11/115102} {\bibfield  {journal} {\bibinfo  {journal}
  {J. Phys. G}\ }\textbf {\bibinfo {volume} {35}},\ \bibinfo {pages} {115102}
  (\bibinfo {year} {2008})},\ \Eprint {http://arxiv.org/abs/0807.3120}
  {arXiv:0807.3120 [hep-ph]} \BibitemShut {NoStop}%
\bibitem [{\citenamefont {Alford}\ and\ \citenamefont
  {Harris}(2018)}]{Alford:2018lhf}%
  \BibitemOpen
  \bibfield  {author} {\bibinfo {author} {\bibfnamefont {M.~G.}\ \bibnamefont
  {Alford}}\ and\ \bibinfo {author} {\bibfnamefont {S.~P.}\ \bibnamefont
  {Harris}},\ }\href {\doibase 10.1103/PhysRevC.98.065806} {\bibfield
  {journal} {\bibinfo  {journal} {Phys. Rev. C}\ }\textbf {\bibinfo {volume}
  {98}},\ \bibinfo {pages} {065806} (\bibinfo {year} {2018})},\ \Eprint
  {http://arxiv.org/abs/1803.00662} {arXiv:1803.00662 [nucl-th]} \BibitemShut
  {NoStop}%
\bibitem [{\citenamefont {Alford}\ and\ \citenamefont
  {Harris}(2019)}]{Alford:2019qtm}%
  \BibitemOpen
  \bibfield  {author} {\bibinfo {author} {\bibfnamefont {M.~G.}\ \bibnamefont
  {Alford}}\ and\ \bibinfo {author} {\bibfnamefont {S.~P.}\ \bibnamefont
  {Harris}},\ }\href {\doibase 10.1103/PhysRevC.100.035803} {\bibfield
  {journal} {\bibinfo  {journal} {Phys. Rev. C}\ }\textbf {\bibinfo {volume}
  {100}},\ \bibinfo {pages} {035803} (\bibinfo {year} {2019})},\ \Eprint
  {http://arxiv.org/abs/1907.03795} {arXiv:1907.03795 [nucl-th]} \BibitemShut
  {NoStop}%
\bibitem [{\citenamefont {Eckart}(1940)}]{Eckart_1940}%
  \BibitemOpen
  \bibfield  {author} {\bibinfo {author} {\bibfnamefont {C.}~\bibnamefont
  {Eckart}},\ }\href {\doibase 10.1103/PhysRev.58.919} {\bibfield  {journal}
  {\bibinfo  {journal} {Phys. Rev.}\ }\textbf {\bibinfo {volume} {58}},\
  \bibinfo {pages} {919} (\bibinfo {year} {1940})}\BibitemShut {NoStop}%
\bibitem [{\citenamefont {Landau}\ and\ \citenamefont
  {Lifshitz}(1987)}]{Landau_1987}%
  \BibitemOpen
  \bibfield  {author} {\bibinfo {author} {\bibfnamefont {L.~D.}\ \bibnamefont
  {Landau}}\ and\ \bibinfo {author} {\bibfnamefont {E.~M.}\ \bibnamefont
  {Lifshitz}},\ }\href@noop {} {\emph {\bibinfo {title} {Fluid Mechanics}}},\
  \bibinfo {edition} {2nd}\ ed.\ (\bibinfo  {publisher}
  {Butterworth-Heinemann},\ \bibinfo {address} {Oxford, England},\ \bibinfo
  {year} {1987})\BibitemShut {NoStop}%
\bibitem [{\citenamefont {Hiscock}\ and\ \citenamefont
  {Lindblom}(1985)}]{Hiscock_1985}%
  \BibitemOpen
  \bibfield  {author} {\bibinfo {author} {\bibfnamefont {W.~A.}\ \bibnamefont
  {Hiscock}}\ and\ \bibinfo {author} {\bibfnamefont {L.}~\bibnamefont
  {Lindblom}},\ }\href {\doibase 10.1103/PhysRevD.31.725} {\bibfield  {journal}
  {\bibinfo  {journal} {Phys. Rev. D}\ }\textbf {\bibinfo {volume} {31}},\
  \bibinfo {pages} {725} (\bibinfo {year} {1985})}\BibitemShut {NoStop}%
\bibitem [{\citenamefont {Hiscock}\ and\ \citenamefont
  {Lindblom}(1987)}]{Hiscock_1987}%
  \BibitemOpen
  \bibfield  {author} {\bibinfo {author} {\bibfnamefont {W.~A.}\ \bibnamefont
  {Hiscock}}\ and\ \bibinfo {author} {\bibfnamefont {L.}~\bibnamefont
  {Lindblom}},\ }\href {\doibase 10.1103/PhysRevD.35.3723} {\bibfield
  {journal} {\bibinfo  {journal} {Phys. Rev. D}\ }\textbf {\bibinfo {volume}
  {35}},\ \bibinfo {pages} {3723} (\bibinfo {year} {1987})}\BibitemShut
  {NoStop}%
\bibitem [{\citenamefont {Gavassino}\ \emph {et~al.}(2020)\citenamefont
  {Gavassino}, \citenamefont {Antonelli},\ and\ \citenamefont
  {Haskell}}]{Gavassino_2020}%
  \BibitemOpen
  \bibfield  {author} {\bibinfo {author} {\bibfnamefont {L.}~\bibnamefont
  {Gavassino}}, \bibinfo {author} {\bibfnamefont {M.}~\bibnamefont
  {Antonelli}}, \ and\ \bibinfo {author} {\bibfnamefont {B.}~\bibnamefont
  {Haskell}},\ }\href {\doibase 10.1103/PhysRevD.102.043018} {\bibfield
  {journal} {\bibinfo  {journal} {Phys. Rev. D}\ }\textbf {\bibinfo {volume}
  {102}},\ \bibinfo {pages} {043018} (\bibinfo {year} {2020})},\ \Eprint
  {http://arxiv.org/abs/2006.09843} {arXiv:2006.09843 [gr-qc]} \BibitemShut
  {NoStop}%
\bibitem [{\citenamefont {Oppenheimer}\ and\ \citenamefont
  {Volkoff}(1939)}]{Oppenheimer:1939ne}%
  \BibitemOpen
  \bibfield  {author} {\bibinfo {author} {\bibfnamefont {J.~R.}\ \bibnamefont
  {Oppenheimer}}\ and\ \bibinfo {author} {\bibfnamefont {G.~M.}\ \bibnamefont
  {Volkoff}},\ }\href {\doibase 10.1103/PhysRev.55.374} {\bibfield  {journal}
  {\bibinfo  {journal} {Phys. Rev.}\ }\textbf {\bibinfo {volume} {55}},\
  \bibinfo {pages} {374} (\bibinfo {year} {1939})}\BibitemShut {NoStop}%
\bibitem [{\citenamefont {{Tolman}}(1939)}]{1939PhRv...55..364T}%
  \BibitemOpen
  \bibfield  {author} {\bibinfo {author} {\bibfnamefont {R.~C.}\ \bibnamefont
  {{Tolman}}},\ }\href {\doibase 10.1103/PhysRev.55.364} {\bibfield  {journal}
  {\bibinfo  {journal} {Phys. Rev.}\ }\textbf {\bibinfo {volume} {55}},\
  \bibinfo {pages} {364} (\bibinfo {year} {1939})}\BibitemShut {NoStop}%
\bibitem [{\citenamefont {Moler}\ and\ \citenamefont {Stewart}(1973)}]{QZ}%
  \BibitemOpen
  \bibfield  {author} {\bibinfo {author} {\bibfnamefont {C.~B.}\ \bibnamefont
  {Moler}}\ and\ \bibinfo {author} {\bibfnamefont {G.~W.}\ \bibnamefont
  {Stewart}},\ }\href {\doibase 10.1137/0710024} {\bibfield  {journal}
  {\bibinfo  {journal} {SIAM J. Numer. Anal.}\ }\textbf {\bibinfo {volume}
  {10}},\ \bibinfo {pages} {241} (\bibinfo {year} {1973})}\BibitemShut
  {NoStop}%
\bibitem [{\citenamefont {Bea}\ and\ \citenamefont {Figueras}(2024)}]{Bea2023}%
  \BibitemOpen
  \bibfield  {author} {\bibinfo {author} {\bibfnamefont {Y.}~\bibnamefont
  {Bea}}\ and\ \bibinfo {author} {\bibfnamefont {P.}~\bibnamefont {Figueras}},\
  }\href {\doibase 10.1007/JHEP11(2024)110} {\bibfield  {journal} {\bibinfo
  {journal} {JHEP}\ }\textbf {\bibinfo {volume} {11}},\ \bibinfo {pages} {110}
  (\bibinfo {year} {2024})},\ \Eprint {http://arxiv.org/abs/2312.16671}
  {arXiv:2312.16671 [hep-th]} \BibitemShut {NoStop}%
\bibitem [{\citenamefont {Kokkotas}\ and\ \citenamefont
  {Ruoff}(2001)}]{Kokkotas:2000up}%
  \BibitemOpen
  \bibfield  {author} {\bibinfo {author} {\bibfnamefont {K.~D.}\ \bibnamefont
  {Kokkotas}}\ and\ \bibinfo {author} {\bibfnamefont {J.}~\bibnamefont
  {Ruoff}},\ }\href {\doibase 10.1051/0004-6361:20000216} {\bibfield  {journal}
  {\bibinfo  {journal} {Astron. Astrophys.}\ }\textbf {\bibinfo {volume}
  {366}},\ \bibinfo {pages} {565} (\bibinfo {year} {2001})},\ \Eprint
  {http://arxiv.org/abs/gr-qc/0011093} {arXiv:gr-qc/0011093} \BibitemShut
  {NoStop}%
\bibitem [{\citenamefont {{Bardeen}}\ \emph {et~al.}(1966)\citenamefont
  {{Bardeen}}, \citenamefont {{Thorne}},\ and\ \citenamefont
  {{Meltzer}}}]{Bardeen:1966}%
  \BibitemOpen
  \bibfield  {author} {\bibinfo {author} {\bibfnamefont {J.~M.}\ \bibnamefont
  {{Bardeen}}}, \bibinfo {author} {\bibfnamefont {K.~S.}\ \bibnamefont
  {{Thorne}}}, \ and\ \bibinfo {author} {\bibfnamefont {D.~W.}\ \bibnamefont
  {{Meltzer}}},\ }\href {\doibase 10.1086/148791} {\bibfield  {journal}
  {\bibinfo  {journal} {\apj}\ }\textbf {\bibinfo {volume} {145}},\ \bibinfo
  {pages} {505} (\bibinfo {year} {1966})}\BibitemShut {NoStop}%
\bibitem [{\citenamefont {Noronha}\ \emph {et~al.}(2022)\citenamefont
  {Noronha}, \citenamefont {Spali{\'n}ski},\ and\ \citenamefont
  {Speranza}}]{Noronha:2021syv}%
  \BibitemOpen
  \bibfield  {author} {\bibinfo {author} {\bibfnamefont {J.}~\bibnamefont
  {Noronha}}, \bibinfo {author} {\bibfnamefont {M.}~\bibnamefont
  {Spali{\'n}ski}}, \ and\ \bibinfo {author} {\bibfnamefont {E.}~\bibnamefont
  {Speranza}},\ }\href {\doibase 10.1103/PhysRevLett.128.252302} {\bibfield
  {journal} {\bibinfo  {journal} {Phys. Rev. Lett.}\ }\textbf {\bibinfo
  {volume} {128}},\ \bibinfo {pages} {252302} (\bibinfo {year} {2022})},\
  \Eprint {http://arxiv.org/abs/2105.01034} {arXiv:2105.01034 [nucl-th]}
  \BibitemShut {NoStop}%
\bibitem [{\citenamefont {Gavassino}(2022)}]{Gavassio_2022:1}%
  \BibitemOpen
  \bibfield  {author} {\bibinfo {author} {\bibfnamefont {L.}~\bibnamefont
  {Gavassino}},\ }\href {\doibase 10.1103/PhysRevX.12.041001} {\bibfield
  {journal} {\bibinfo  {journal} {Phys. Rev. X}\ }\textbf {\bibinfo {volume}
  {12}},\ \bibinfo {pages} {041001} (\bibinfo {year} {2022})},\ \Eprint
  {http://arxiv.org/abs/2111.05254} {arXiv:2111.05254 [gr-qc]} \BibitemShut
  {NoStop}%
\bibitem [{\citenamefont {Gavassino}\ \emph {et~al.}(2022)\citenamefont
  {Gavassino}, \citenamefont {Antonelli},\ and\ \citenamefont
  {Haskell}}]{Gavassino_2022:2}%
  \BibitemOpen
  \bibfield  {author} {\bibinfo {author} {\bibfnamefont {L.}~\bibnamefont
  {Gavassino}}, \bibinfo {author} {\bibfnamefont {M.}~\bibnamefont
  {Antonelli}}, \ and\ \bibinfo {author} {\bibfnamefont {B.}~\bibnamefont
  {Haskell}},\ }\href {\doibase 10.1103/PhysRevLett.128.010606} {\bibfield
  {journal} {\bibinfo  {journal} {Phys. Rev. Lett.}\ }\textbf {\bibinfo
  {volume} {128}},\ \bibinfo {pages} {010606} (\bibinfo {year} {2022})},\
  \Eprint {http://arxiv.org/abs/2105.14621} {arXiv:2105.14621 [gr-qc]}
  \BibitemShut {NoStop}%
\bibitem [{\citenamefont {Lier}\ \emph {et~al.}(2026)\citenamefont {Lier},
  \citenamefont {Armas},\ and\ \citenamefont {Porth}}]{Lier:2026owd}%
  \BibitemOpen
  \bibfield  {author} {\bibinfo {author} {\bibfnamefont {R.}~\bibnamefont
  {Lier}}, \bibinfo {author} {\bibfnamefont {J.}~\bibnamefont {Armas}}, \ and\
  \bibinfo {author} {\bibfnamefont {O.}~\bibnamefont {Porth}},\ }\href@noop {}
  {\  (\bibinfo {year} {2026})},\ \Eprint {http://arxiv.org/abs/2606.22691}
  {arXiv:2606.22691 [astro-ph.HE]} \BibitemShut {NoStop}%
\bibitem [{\citenamefont {Most}\ and\ \citenamefont
  {Dunham}(2026)}]{Most:2026uga}%
  \BibitemOpen
  \bibfield  {author} {\bibinfo {author} {\bibfnamefont {E.~R.}\ \bibnamefont
  {Most}}\ and\ \bibinfo {author} {\bibfnamefont {S.~J.}\ \bibnamefont
  {Dunham}},\ }\href@noop {} {\  (\bibinfo {year} {2026})},\ \Eprint
  {http://arxiv.org/abs/2609.21923} {arXiv:2609.21923 [astro-ph.HE]}
  \BibitemShut {NoStop}%
\bibitem [{\citenamefont
  {\rm{Information~Systems~and~Wake~Forest~University}}(2021)}]{WakeHPC}%
  \BibitemOpen
  \bibfield  {author} {\bibinfo {author} {\bibnamefont
  {\rm{Information~Systems~and~Wake~Forest~University}}},\ }\href {\doibase
  10.57682/G13Z-2362} {\enquote {\bibinfo {title} {{WFU High Performance
  Computing Facility}},}\ } (\bibinfo {year} {2021})\BibitemShut {NoStop}%
\end{thebibliography}%
\end{document}